\documentclass[pre,superscriptaddress]{revtex4}
\usepackage{amsfonts}
\usepackage{amsmath}
\usepackage{amssymb}
\usepackage{graphicx}

\newcommand{\beq}{\begin{equation}}
\newcommand{\eeq}{\end{equation}}
\newcommand{\be}{\begin{eqnarray}}
\newcommand{\ee}{\end{eqnarray}}
\newcommand{\bk}{{\bf k}}
\newcommand{\bq}{{\bf q}}
\newcommand{\bg}{{\bf g}}
\newcommand{\bx}{{\bf x}}
\newcommand{\br}{{\bf r}}
\newcommand{\bR}{{\bf R}}
\newcommand{\bS}{{\bf S}}

\begin{document}

\title{Recent developments of the Hierarchical Reference Theory of Fluids and 
its relation to the Renormalization Group}
\author{Alberto Parola}
\affiliation{Dipartimento di Scienza e Alta Tecnologia, Universit\`a dell'Insubria,
Via Valleggio 11, 22100 Como, Italy}
\author{Luciano Reatto}
\affiliation{Dipartimento di Fisica, Universit\`a degli Studi di Milano,
Via Celoria 16, 20133 Milano, Italy}

%\pacs{64.60.F-, 61.20.Gy, 64.60.A-, 05.70.Fh}

%%%%%%%%%%%%%%%%%%%%%%%%%%%%%%%%%%%%%%%%%%%%%%%%%%%%%%%%%%%%%%%%%%%%%%%
\begin{abstract}
The Hierarchical Reference Theory (HRT) of fluids is a general framework
for the description of phase transitions in microscopic models of classical and
quantum statistical physics. The foundations of HRT are briefly reviewed 
in a self-consistent formulation which includes both the original sharp cut-off 
procedure and the smooth cut-off implementation, which has been recently investigated.
The critical properties of HRT are summarized, together 
with the behavior of the theory at first order phase transitions. 
However, the emphasis of this presentation 
is on the close relationship between HRT and non perturbative renormalization
group methods, as well as on recent generalizations of HRT to microscopic models 
of interest in soft matter and quantum many body physics. 
\end{abstract}
\maketitle
%%%%%%%%%%%%%%%%%%%%%%%%%%%%%%%%%%%%%%%%%%%%%%%%%%%%%%%%%%%%%%%%%%%%%%%%
\section{Introduction}

Liquid State Theory experienced remarkable successes in the understanding of 
the properties of dense, simple fluids
\cite{hansen} and more recently, of complex liquids, like colloidal suspensions or 
globular polymers \cite{barrat}. 
However a fully quantitative description of phase transitions cannot be obtained by 
standard integral equations
or diagrammatic expansions and requires specially devised approaches. The freezing transition, often 
dominated by geometrical constraints, is usually understood on the basis of Density Functional Theory
\cite{hansen}, while 
fluctuation induced phenomena, like the liquid-vapour transition, need the completely different theoretical 
framework provided by the Renormalization Group method \cite{rg}.

The Hierarchical Reference Theory of Fluids (HRT) was formulated more than 25 years ago 
as an attempt to reconcile the Renormalization Group (RG) approach, an extremely 
general and successful theory of second order phase transitions formulated in
field-theoretical language, and the
microscopic point of view adopted within Liquid State Theory \cite{hrt0}. 
The main goal of HRT was to develop a unified theory able to reproduce the
universal properties at the critical point, as obtained by RG techniques, together 
with the non universal features of realistic models of fluids, 
present both in the critical region (e.g. critical density and temperature, short range 
correlations) as well as far from the critical point. 

A comprehensive presentation of the HRT formalism, 
including a detailed derivation of the basic equations and an extensive discussion of
its critical properties, can be found in Ref. \cite{adv}. The earliest applications of HRT
to simple fluids and to the Ising model are briefly discussed in the same reference. However,
most of the more interesting results and extensions were developed later.
By now it is therefore appropriate to summarize the advances in HRT 
which took place in the last decade, providing a guidance to the most 
significant recent applications. 

Such a critical discussion of the
HRT formalism also gives the opportunity to emphasize its relationship with other 
theoretical tools which have been independently developed in different frameworks.
Since the early nineties, 
sophisticated functional, non perturbative RG techniques (NPRG) were devised within
field theory, leading to a boost of applications
to many and diverse physical situations: from particle physics, to quantum many body theory, to
classical statistical mechanics \cite{wetterich}. Only recently it was recognized that,
despite the different language and physical context, a deep relation ties HRT and NPRG \cite{caillol},
whose developments have proceeded quite independently for 20 years 
in distinct scientific communities.

\section{The HRT for classical fluids}
\label{sechrt}

To introduce the Hierarchical Reference Theory of Fluids it is convenient to refer 
to a system of $N$ classical particles confined 
in a volume $V$, in spatial dimension $d$, interacting via a pairwise
potential $v(\br)$. Generalizations to classical spin systems, quantum fluids and magnets do 
not present additional difficulties and will be examined in the next Sections. 
The grand canonical partition function $\Xi$ of the model is written as: 
\begin{equation} 
\Xi[u(\br)]= \sum_N \frac{z^{N}}{N!}\int {\rm d}{ \bf r_1}...{\rm d}{ \bf r_N} \,
e^{ -\beta\sum_{i<j} v({\bf r}_i - {\bf r}_j)-\beta\sum_i u({\bf r}_i)}
\label{gran}
\end{equation}
where the fugacity $z$ is related to the chemical potential $\mu$ and 
to the thermal wavelength $\Lambda=h/\sqrt{2\pi m k_B T}$  by $z=\Lambda^{-d}\,e^{\beta\mu}$ with 
$\beta=(k_B T)^{-1}$. An external field $u(\br)$ is also included for convenience and
$\Xi$ is considered as a functional of $u(\br)$. 
A detailed derivation of the HRT equations and a thorough discussion of their properties
can be found in Ref. \cite{adv} via diagrammatic methods \cite{note1}. 
Here we provide a brief introduction of the physical 
content of the method together with a simpler, less technical derivation. 

Since the pioneering works by Van der Waals, a large part of Liquid State 
Theory is based on the idea that the strongly repulsive short range interaction,
distinctive of the interatomic potentials, provides a constraint to the 
available phase space, determining an important contribution to the entropy of the fluid, 
while the mainly attractive, regular part of $v(\br)$ dominates the internal energy. 
This is certainly true for the special, although prototypical, case 
in which $v(\br)$ has  hard core at short distances, but holds also
in the general case of soft core potentials. The different role played by 
the singular and regular contributions to the interparticle potential is 
manifest in lattice gas models, where particles live on the sites of a lattice: 
hard core repulsion between particles gives rise to the single occupancy constraint, 
which allows to map the lattice gas onto an Ising model with a spin-spin 
interaction proportional to the regular part of the interparticle potential.
Mean field approaches developed for 
specific models, like Van der Waals equation in simple fluids or Weiss theory in 
magnets, suggest that only the latter interaction triggers the order parameter fluctuations,
responsible for the order-disorder transitions.

In fluid models, both the singular and the regular part contributions are included in
the interaction $v(\br)$ which should then be split in a reference, singular, part $v_R(\br)$
and in the regular remainder $w(\br)=v(\br)-v_R(\br)$. 
The partition function (\ref{gran}) can be written in a suggestive form by introducing the 
``configurational density" $\hat\rho(\br)=\sum_i \delta(\br-\br_i)$ whose 
thermal average is just the density profile $\rho(\br)$, which reduces to the 
ordinary average density $\rho$ for $u(\br)=0$. 
In the following, the properties of the reference system will be considered as known.
The fluctuating order parameter of the liquid-vapour phase transition
in fluids is then identified as the configurational density and $\Xi$ is written as
\begin{equation}
\Xi[u(\br)] = \sum_N\frac{(z\,e^{-\frac{1}{2}\phi(0)})^{N}}{N!}\, 
\int {\rm d}{ \bf r_1}...{\rm d}{ \bf r_N} \, e^{-\beta H_R}\,
e^{\frac{1}{2}\int {\rm d}\bx_1 \,{\rm d}\bx_2 \phi(\bx_1-\bx_2) \, \hat\rho(\bx_1)
\,\hat\rho(\bx_2)}
\label{xi1}
\end{equation}
where the regular dimensionless interaction has been defined as $\phi(\br)=-\beta\,w(\br)$ 
and the reference hamiltonian is
\begin{equation}
H_R = \sum_{i<j} v_R({\bf r}_i - {\bf r}_j)+ \int {\rm d}\bx \, u(\bx)\hat\rho(\bx)
\end{equation}
According to the previous remarks, $H_R$ 
defines a non-trivial integration measure, limiting the phase space available 
to the particles. By factorizing the reference $N$-particle canonical partition function 
$Z_R=e^{-\beta\,A_R(N)}$, 
Eq. (\ref{xi1}) can be also expressed as: 
\begin{equation}
\Xi[u(\br)] = 
\sum_N\frac{(z\,e^{-\frac{1}{2}\phi(0)})^{N}}{N!}\,e^{-\beta\,A_R(N)} \,\left \langle  
\exp\left (\frac{1}{2}\int \frac{{\rm d}\bq}{(2\pi)^d} \,\tilde\phi(\bq)\,
\hat\rho(\bq)\,\hat\rho(-\bq)\right ) \right \rangle_R
\label{xi2}
\end{equation}
where the average $\langle \cdots\rangle_R$ refers to the Gibbs measure of the
reference fluid, and the Fourier transforms of the potential $\tilde\phi(\bq)$ and of the
density fluctuations $\hat\rho(\bq)=\sum_n e^{i\bq\cdot\br_n}$  have been defined. 

Such a form is particularly suitable as a starting point for a reformulation of 
Wilson's Renormalization Group picture in hamiltonian systems. In the momentum space RG 
formulation \cite{rg,delamotte}, a sequence of systems, labeled by a wave-vector $Q$, is defined
by allowing density fluctuations only at wave-vectors $\bq$ larger than 
the cut-off $Q$. The evolution of the physical properties of the model is then followed 
starting from the limit 
$Q=\infty$, where the system reduces to the reference one with a 
shift $w(0)/2$ in the chemical potential, to the final value
$Q=0$, where the cut-off is removed and the original partition function $\Xi$ is retrieved. 
Expression (\ref{xi2}) can easily accommodate the presence of a cut-off $Q$: 
\begin{equation}
\Xi_Q[u(\br)] = 
\sum_N\frac{(z\,e^{-\frac{1}{2}\phi(0)})^{N}}{N!}\,e^{-\beta\,A_R(N)} \,\left \langle  
\exp\left (\frac{1}{2}\int \frac{{\rm d}\bq}{(2\pi)^d} \,\tilde\phi(\bq)\,\theta (q-Q)\,
\hat\rho(\bq)\,\hat\rho(-\bq)\right ) \right \rangle_R
\label{xiq}
\end{equation} 
where $\theta(x)$ is the Heaviside step function. 
This definition embodies basic principle of HRT: A cut-off in density 
fluctuations is equivalent to a cut-off in the regular part of the two body 
interaction for the evaluation of the fluctuation corrections to the free energy, 
leading to the definition of a cut-off dependent potential
$\tilde w_Q(\bq) = \tilde w(\bq)\,\theta(q-Q)$. 
It is customary to perform a Legendre transform defining a cut-off dependent Helmholtz free 
energy $A_Q$ as a functional of the density profile $\rho(\br)$: 
\begin{equation}
A_Q[\rho(\br)] = -kT\,\log \Xi_Q[u(\br)] - \int {\rm d}\br\,
\left (u(\br)-kT\,\log z\right )\,\rho(\br)
\label{leg0}
\end{equation}
Finally, it is also convenient to include in the definition of the free energy functional 
the mean field contribution due to the residual interaction $w(\br)-w_Q(\br)$: 
\begin{equation}
{\cal A}_Q[\rho(\br)] = A_Q[\rho(\br)] 
+ \frac{1}{2}\int {\rm d}\br \,{\rm d}\br^\prime
\left [ w(\br-\br^\prime) - w_Q(\br-\br^\prime)\right ] \, \rho(\br) \,\rho(\br^\prime)
\label{leg}
\end{equation}
In this way, for $Q\to\infty$, ${\cal A}_Q$ reduces to the mean field 
approximation of the full free energy, which is recovered as $Q\to 0$.
This way to artificially 
eliminate the long wave-length fluctuations from the thermodynamics is 
referred to as ``sharp cut-off". However it is often more convenient to deal with a ``smooth cut-off" 
procedure, where the $\theta$-function in the definition of $\tilde w_Q(\bq)$ 
is substituted by a 
smoothed step function, denoted in the following as $\theta_\epsilon(x)$ which 
is continuous and differentiable for every $\epsilon > 0$ and tends to the step function
in the $\epsilon \to 0$ limit: 
\begin{equation}
\tilde w_Q(\bq) = \tilde w(\bq)\,\theta_\epsilon(q-Q) 
\label{wqsmooth}
\end{equation}

\subsection{The exact hierarchy in smooth and sharp cut-off formulations}
\label{sec:exact}

After having defined the cut-off dependent 
free energy, ${\cal A}_Q$ via Eq. (\ref{leg}), the next step 
is to write the exact evolution (or flow) equation governing the change in the physical properties
of the model when the cut-off $Q$ is varied. This goal is most easily achieved in the smooth
cut-off formulation, where an infinitesimal increase of the cut-off $Q$ leads to an 
infinitesimal change in the potential $\tilde w_Q(\bq)$. 
According to standard first order perturbation theory \cite{hansen} we get, in a uniform fluid, 
\begin{eqnarray}
\label{evol1}
\frac{1}{V}
\frac{\partial (-\beta {\cal A}_Q) }{\partial Q} &=& \frac{\rho^2}{2}\,\int \,{\rm d}\br \,h_Q(\br)\, 
\frac{\partial \phi_Q(\br)}{\partial Q} + \frac{\rho}{2} \frac{\partial\phi_Q(0)}{\partial Q} \\
&=& \frac{1}{2}\,\int \,{\rm d}\br \,F_Q(\br)\, \frac{\partial \phi_Q(\br)}{\partial Q}
\label{smooth0}
\end{eqnarray}
where $h_Q(\br)=g_Q(\br)-1$ is the total correlation function of a fluid interacting 
through a potential $w_Q(\br)$ and $F_Q(\br) = \rho\, \delta (\br) + \rho^2 h_Q(\br)$.
This smooth cut-off HRT equation has been obtained for the first time in Ref. \cite{ap}
as an attempt to apply the Renormalization Group ideas to microscopic models. 
Note that the total correlation function $h_Q$ rather than
the radial distribution function $g_Q$ appears in the evolution equation (\ref{evol1}) 
because the mean field contribution 
has been subtracted in the definition of the $Q$-dependent free energy (\ref{leg}). 
The total correlation function can be conveniently expressed in terms of the direct correlation
function $c_Q(\br)$ by use of the Ornstein-Zernike relation:
\begin{equation}
\tilde h_Q(\bq) = \frac{\tilde c_Q(\bq)}{1-\rho \,\tilde c_Q(\bq)}
\label{oz}
\end{equation}
The limit $\epsilon\to 0$ in Eq. (\ref{wqsmooth}) requires some care, as already 
recognized in the description of the sharp cut-off HRT equation and also discussed
in field theoretical approaches \cite{morris}, because of the emergence of a discontinuity in
$h_Q(\bq)$ precisely on the same surface $q=Q$ where the $\delta$-function singularity 
develops in $\partial_Q \tilde w_Q(\bq)$. Such a discontinuity originates from the 
Random Phase contribution to the exact direct correlation function, 
being the single term where the regular part of the potential appears outside any integration.
If $\tilde c_R(\bq)$ denotes the direct correlation function of the reference fluid,
the first terms of the perturbative expansion read
\begin{equation}
\tilde c_Q(\bq) = \tilde c_R(\bq) + \tilde \phi_Q(\bq) + \cdots
\end{equation}
All the remaining contributions, being integrated, do not display discontinuities in the whole
$\bq$-space. This suggests to introduce a modified correlation function ${\cal C}_Q(\bq)$ 
which remains continuous also in the sharp cut-off limit $\epsilon\to 0$:
\begin{equation}
{\cal C}_Q(\bq) = \tilde c_Q(\bq) - \tilde \phi_Q(\bq) + \tilde \phi(\bq) 
\label{calc}
\end{equation}
Interestingly, this newly defined function is just the second functional derivative of
the free energy functional $-\beta {\cal A}_Q[\rho(\br)]$ with respect to density variations, 
evaluated at uniform density: 
\begin{equation}
\frac{\delta^2 (-\beta {\cal A}_Q)} {\delta \rho(\br_1)\,\delta \rho(\br_2)} \Big\vert_{\rho(\br)=\rho}
= {\cal C}_Q(\br_1-\br_2) - \frac{\delta(\br_1-\br_2)}{\rho}
\label{second}
\end{equation}
where the last term is due to the ideal gas contribution, included in our definition of ${\cal A}_Q$.
We first write the right hand side of Eq. (\ref{evol1}) by introducing the Fourier transforms:
\begin{equation}
\frac{1}{V}
\frac{\partial (-\beta {\cal A}_Q)}{\partial Q} = \frac{\rho}{2}\,\int \,\frac{{\rm d}\bq}{(2\pi)^d} \, 
\frac{1}{1 - \rho\,{\cal C}_Q(\bq) +\rho\,\left [\tilde \phi(\bq) - \tilde \phi_Q(\bq)\right ]}
\, \frac{\partial \tilde\phi_Q(\bq)}{\partial Q}
\label{smooth}
\end{equation}
The easiest way to take the sharp cut-off limit of this flow equation 
is to introduce the differential operator $\bar \partial_Q$ acting only on the 
Q-dependence of $\tilde\phi_Q(\bq)$ (while ${\cal C}_Q(\bq)$ is treated as $Q$-independent):
\begin{equation}
\frac{1}{V} \frac{\partial (-\beta {\cal A}_Q)}{\partial Q} 
= - \frac{1}{2}\,\int \,\frac{{\rm d}\bq}{(2\pi)^d} \,
\bar \partial_Q \log \left [ 1 - \rho\,\left ( {\cal C}_Q(\bq) +
\tilde \phi_Q(\bq) - \tilde \phi(\bq)\right ) \right ] 
\label{evol2}
\end{equation}
Taking the $\epsilon\to 0$ limit and 
writing $\bar\partial_Q$ as a finite difference, Eq. (\ref{evol2}) is then written in the 
form \cite{hrt0}:
\begin{equation}
\frac{1}{V}
\frac{\partial (-\beta {\cal A}_Q)}{\partial Q} = - \frac{1}{2}\,\int_{q=Q} \,
\frac{{\rm d}\Omega_\bq}{(2\pi)^d} 
\, \log \left [ 1 + {\cal F}_Q(\bq)\tilde\phi(\bq)\right ]
\label{sharp}
\end{equation}
where the integration is limited to the spherical surface (in dimension $d$) defined by $q=Q$
and the function ${\cal F}_Q(\bq)$ is defined by the Ornstein-Zernike relation as:
\begin{equation}
{\cal F}_Q(\bq)=\frac{\rho}{1-\rho \,{\cal C}_Q(\bq)}
\label{calf}
\end{equation}
Equations (\ref{smooth}) and (\ref{sharp}) govern the HRT flow of the free energy 
in the smooth and sharp formulations respectively. As previously mentioned, the initial
condition at $Q=\infty$ coincides in both cases with the known mean field approximation, while
for $Q\to 0$ all density fluctuations are included and the exact free energy is in principle recovered
if ${\cal F}_Q(\bq)$ is known.  

HRT then provides the exact evolution of the free energy in terms of the 
two body correlation function $F_Q(\bq)$ (or equivalently ${\cal F}_Q(\bq)$) whose flow is determined
by similar equations involving the three and four particle correlations \cite{hrt0,ap}.
These equations, which can be derived along the same lines previously sketched, are 
most easily obtained by functional differentiation of (\ref{smooth}) and (\ref{sharp})
with respect to $\rho(\br)$ leading to a hierarchy of differential equations
describing the RG flow of the free energy and of the $n$-particle direct correlation functions $c_n^Q(\br_1\cdots\br_n)$,
formally defined as the $n^{th}$ functional derivative of $(-\beta {\cal A}_Q)$ with respect to the density profile
$\rho(\br)$ evaluated at uniform density $\rho(\br)=\rho$. The ideal gas contribution
is included in our definition of ${\cal A}_Q$ leading to an additional term
$(n-2)!\,(-\rho)^{1-n} \prod_{i\ne 1} \,\delta(\br_i-\br_1)$ with respect to the standard
definition of the many particle direct correlation functions \cite{adv}.

As an example, we explicitly report the flow equations for the previously defined two point function
in smooth cut-off formulation \cite{ap}
\begin{eqnarray}
\frac{\partial\tilde c_Q(\bk)}{\partial Q} &=&  \frac{1}{2} \int \,\frac{{\rm d}\bq}{(2\pi)^d} \,
c_4^Q(\bk,-\bk,\bq,-\bq) \, F_Q^2(\bq) \, \frac{\partial \tilde\phi_Q(\bq)}{\partial Q} +\nonumber \\
&& \int \,\frac{{\rm d}\bq}{(2\pi)^d} \,
c_3^Q(\bk,\bq,-\bk-\bq)\,F_Q(\bk+\bq)\,c_3^Q(-\bk,-\bq,\bk+\bq)\, F_Q^2(\bq)\, \frac{\partial \tilde\phi_Q(\bq)}{\partial Q}
\label{csmooth}
\end{eqnarray}
and in sharp cut-off formulation \cite{hrt0}:
\begin{eqnarray}
-\frac{\partial {\cal C}_Q(\bk)}{\partial Q} &=& 
\frac{1}{2} \int_{q=Q} \,\frac{{\rm d}\Omega_\bq}{(2\pi)^d} \,
c_4^Q(\bk,-\bk,\bq,-\bq)\, \frac{{\cal F}_Q^2(\bq)\tilde\phi(\bq)}{1+{\cal F}_Q(\bq)\tilde\phi(\bq)} 
+\nonumber \\
&& \int_{q=Q} \,\frac{{\rm d}\Omega_\bq}{(2\pi)^d} \,
c_3^Q(\bk,\bq,-\bk-\bq) \, F_Q(\bk+\bq)\,c_3^Q(-\bk,-\bq,\bk+\bq) 
\frac{{\cal F}_Q^2(\bq)\tilde\phi(\bq)}{1+{\cal F}_Q(\bq)\tilde\phi(\bq)}
\label{csharp}
\end{eqnarray}
where, in the latter equation, $F_Q(\bk+\bq)$ is defined as:
\begin{equation}
F_Q(\bk+\bq) = 
\begin{cases}
{\cal F}_Q(\bk+\bq)\,\left [1+{\cal F}_Q(\bk+\bq)\tilde\phi(\bk+\bq)\right ]^{-1} & \text{for $\vert \bk+\bq\vert < Q$} \\
{\cal F}_Q(\bk+\bq) & \text{for $\vert \bk+\bq\vert > Q$} 
\end{cases}
\end{equation}
Note that, in the limit $\bk\to 0$, these flow equations reproduce exactly the second density derivative of the 
corresponding free energy evolution (\ref{smooth},\ref{sharp}), 
showing that the density dependence of the free energy flow encodes some (partial) information 
contained in the full infinite hierarchy. 

\subsection{HRT as a microscopic implementation of the Functional RG}
\label{sec:nprg}

Since the first derivation of the HRT equations in the sharp cut-off formulation \cite{hrt0} a 
close similarity to some previous differential RG formulation was noticed. In particular it was shown
that close to a critical point, in the long wave-length limit the full HRT hierarchy 
precisely reproduces 
the flow equations for the one particle irreducible RG generator \cite{nicoll}. The analogy 
between the Hierarchical Reference Theory and functional RG approaches is however even deeper,
as recently shown by Caillol \cite{caillol}. To highlight the connection between the two approaches
it is convenient to start from the definition of the partition function $\Xi_Q$ 
(\ref{xiq}). As pointed out in an early attempt to apply RG techniques to microscopic models \cite{hubbard}, by
performing a Hubbard-Stratonovich transformation, the partition function can be 
expressed as a functional integral over an auxiliary scalar field $\psi_\bq$ 
governed by an effective action $S_Q[\psi_\bq]$. This commonly adopted procedure, however, 
is rather formal and does not lend a well defined physical meaning for
the Hubbard Stratonovich field $\psi_\bq$.
It is possible to circumvent this problem by first introducing a reference effective
action $S_R[\psi_\bq]$ by:
\begin{equation}
S_R[\psi_\bq] = -\log \left [ 
\sum_N\frac{z^{N}}{N!}\, 
\int {\rm d}{ \bf r_1}...{\rm d}{ \bf r_N} \, e^{-\beta\sum_{i<j} v_R({\bf r}_i - {\bf r}_j)}\,
\prod_\bq \delta(\hat\rho(\bq)-\psi_\bq)
\right ] 
\label{sr}
\end{equation}
This definition clearly forces the effective field $\psi_\bq$ to 
faithfully represent density fluctuations. 
The exact partition function is then written as 
\begin{eqnarray}
\label{func}
\Xi_Q[u(\br)] &=& \int {\cal D} \psi \, e^{-S_Q[\psi_\bq]} \\
S_Q[\psi_\bq] &=& S_R[\psi_\bq] - \frac{1}{2V}\,\sum_\bq \tilde\phi_Q(\bq)\,
\psi_\bq\,\psi_{-\bq} + \frac{1}{V} \sum_\bq \left ( \frac{\phi(0)}{2} + \beta \tilde u(\bq)\right )
\psi_{-\bq} 
\label{action}
\end{eqnarray}
This expression shows that $S_Q[\psi_\bq]$ can be written as the 
sum of the effective action in the absence of any cut-off $S[\psi_\bq]$ (i.e. 
the value attained at $Q=0$) plus a ``momentum dependent mass term" \cite{wetterich}: 
\begin{equation}
S_Q[\psi_\bq] = S[\psi_\bq] + \frac{1}{2V}\,\sum_\bq R_Q(\bq) \, \psi_\bq\,\psi_{-\bq}
\end{equation}
with 
\begin{equation}
R_Q(\bq) = \tilde\phi(\bq) - \tilde\phi_Q(\bq) = \tilde\phi(\bq) \,
\left [ 1-\theta_\epsilon(q-Q)\right] 
\label{mass}
\end{equation} 
Following Wetterich we now perform a ``modified Legendre transform" \cite{wetterich,delamotte}
introducing the field $\rho_\bq$ conjugated to the external potential $\tilde u(\bq)$, which, due to the
definition (\ref{action}) coincides with the statistical average of the field $\psi_\bq$:
\begin{equation}
\Gamma[\rho_\bq] = -\log\Xi_Q + \frac{1}{V}\,\sum_\bq 
\left ( \tilde u(\bq)-kT\,\log z\right )\,\rho_{-\bq}- 
\frac{1}{2V}\,\sum_\bq R_Q(\bq)\, \rho_\bq\,\rho_{-\bq}
\label{leg1}
\end{equation}
Comparing this definition to Eq. (\ref{leg}) it is clear that the average action 
coincides with the modified free energy defined in HRT: $\Gamma = \beta {\cal A}_Q$,
providing  the link between HRT, based on the gradual inclusion of the 
Fourier components of the regular part of the physical interaction, 
and the ``effective average action" RG
method built upon the idea of adding a cut-off and momentum dependent mass term $R_Q(\bq)$ to the original action. 
From Eq. (\ref{mass}) it follows that 
for $Q=0$ the mass term vanishes identically, while at fixed $Q$, $R_Q(\bq)$ 
is large only for $q \lesssim Q$ providing an effective cut-off for the
long wave-length fluctuations, as requested in all the implementations of the momentum space 
RG ideas \cite{delamotte}. A difference in the initial condition however persists between 
the average action method and HRT:  in the RG approach, $R_Q(\bq)$ 
diverges as $Q\to\infty$, thus providing an infinite mass to all fluctuations and 
leading to a soluble problem. 
Instead in HRT we get $\lim_{Q\to\infty} R_Q(\bq) = \tilde\phi(\bq)$: for this special 
but finite value of the mass term the regular part of the interaction cancels exactly 
in $S_Q[\psi_\bq]$ and the 
model reduces to the soluble reference system. The procedure adopted in HRT is reminiscent of 
some recent development in functional RG methods \cite{dupuis}. 

\subsection{Approximate closures}
\label{appro}

The solution of the infinite hierarchy of equations previously derived needs some approximation scheme. 
Two kinds of approximations are customary in this type of problems: 
truncations and closures. A truncation 
of the hierarchy simply corresponds to neglecting $n$-particle correlations for $n$ 
larger than a maximum value $n^*$
(typically $n^*=4$). Such a procedure has been followed in the early investigations 
of RG equations but it is rigorously
justified only near a critical point in spatial dimension close to $d=4$ \cite{rg}, while 
it is not expected to provide an accurate description of a system, like a liquid, where short range 
correlations are important.
Instead, most of the applications of HRT, both in the smooth and sharp cut-off formulations, 
adopted a non perturbative closure to the first equation of the hierarchy (\ref{smooth},\ref{sharp})
and only this type of approximation will be discussed here. Non perturbative closures 
in fact allow to preserve the convexity of the free energy, providing a physically 
satisfactory description of phase transitions between homogeneous phases, together with 
a picture of the critical region fully consistent with the scaling hypothesis. 

The evolution equation for the free energy involves the knowledge of the two particle direct 
correlation function in the presence of a cut-off $Q$. A closure is a parametrization of
the direct correlation function expressing ${\cal C}_Q(\bk)$ in terms of the free energy density 
$A_Q/V$. As previously noticed, the compressibility relation, which follows directly from Eq. (\ref{second}),
\begin{equation}
\frac{\partial^2 }{\partial\rho^2} \,\left ( \frac{-\beta {\cal A}_Q}{V}\right )
= {\cal C}_Q(\bk=0) - \frac{1}{\rho}
\label{komp}
\end{equation}
plays a special role in HRT because, when implemented into the free energy evolution equation, allows to 
include in the very first equation some information contained into the whole hierarchy.

The simplest non perturbative closure of the first HRT equation 
is inspired by the well known Random Phase Approximation (RPA) in Liquid 
State Theory \cite{hansen}:
\begin{equation}
{\cal C}_Q(\bk) = c_R(\bk) + \lambda_Q \, \phi(\bk)
\label{rpa}
\end{equation}
The exact initial condition at $Q\to\infty$ is obtained for 
 $\lambda_Q=1$, while the compressibility relation (\ref{komp}) can be enforced 
at all $Q$ via a suitable choice of the parameter $\lambda_Q$. 
When this closure is substituted into Eq. (\ref{smooth}) or (\ref{sharp}) the 
resulting partial differential equation for the free energy ${\cal A}_Q$ as a function
of density $\rho$ and cut-off $Q$ can be solved for a given value of the temperature. 

Interestingly, this approximate closure can be equivalently formulated 
starting from the evolution equations
(\ref{smooth}) or (\ref{sharp}) for the full functional ${\cal A}_Q[\rho(\br)]$ \cite{hrt0}:
By restricting its form through a suitably chosen parametrization, 
the functional flow equation gives rise to partial differential
equations for the parameters. For instance, the RPA-like closure is obtained by 
use of a simple Weighted Density Approximation \cite{tarazona} for the free energy:
\begin{equation}
{\cal A}_Q [\rho(\br)] = A_R[\rho(\br)] + \int {\rm d}\br \,f_Q(\bar\rho(\br)) 
\label{wda}
\end{equation}
where $A_R[\br]$ is the exact free energy functional of the reference system and 
the function $f_Q(\rho)$ 
is the parameter whose evolution will be determined by the flow equation.
The weighted density in Eq. (\ref{wda}) $\bar\rho(\br)$ is defined by the local average:
\begin{equation}
\bar\rho(\br) = \int {\rm d}\br^\prime \, \omega(\br-\br^\prime) \, \rho(\br^\prime) 
\label{wda1}
\end{equation}
with a normalized weight function given by 
$\tilde \omega(\bq) = [\tilde\phi(\bq)/\tilde\phi(0)]^{1/2}$ \cite{note3}. 
By substituting this form into the evolution equation (\ref{smooth}) or (\ref{sharp}), 
we obtain the same expression found by use of the ansatz (\ref{rpa}) if we identify 
$\lambda_Q \, \tilde w(0)= \frac{\partial^2 f_Q}{\partial\rho^2}$. This 
class of closures, in the NPRG framework, are known as Local Potential Approximation (LPA)
because of the similarity between Eq. (\ref{wda}) and the usual Local Density Approximation
of Density Functional Theory.

Clearly, more refined approximations to the direct correlation function, or equivalently
to the free energy functional, may be devised, leading to more accurate 
representation of the exact free energy flow. In fluids with a hard sphere 
reference system (in the continuum or on a lattice), Eq. (\ref{rpa}) can be
generalized by including a further contribution, non vanishing only
inside the hard sphere diameter, so as to satisfy the core condition 
$g_Q(r)= 0$. This constraint can be implemented exactly in the smooth cut-off formulation
for Yukawa potentials \cite{smooth} thanks to the exact analytical solution of the 
Ornstein-Zernike equation for this model \cite{msa}. 

Analogous closures might be implemented on the flow equation for the two 
particle direct correlation function 
(\ref{csmooth}) or (\ref{csharp}). A closure to the second equation of
the hierarchy amounts to expressing the three and four particle correlations in 
terms of the two point function. A possible closure has been proposed already 
in the earliest stage of HRT \cite{hrt0,ap} but it has not been implemented yet:
\begin{eqnarray}
\label{c3}
c_3^Q(\bk,\bq,-\bk-\bq) &=& \frac{\partial {\cal C}_Q(\bk+\bq)}{\partial\rho} \\
c_4^Q(\bk,-\bk,\bq,-\bq) &=& 
\frac{1}{2}\left [ \frac{\partial^2 {\cal C}_Q(\bk+\bq)}{\partial\rho^2} +
\frac{\partial^2 {\cal C}_Q(\bk-\bq)}{\partial\rho^2} \right ]
\label{c4}
\end{eqnarray}
This class of approximations 
has the advantage of allowing for a non trivial anomalous dimension exponent
at the critical point \cite{hrt0,ap}. 
A similar closure has been independently proposed in the context
of functional RG \cite{bmw} and a simplified form has been numerically investigated 
in the same framework \cite{bmw2}. 

Up to now, all the applications of HRT to microscopic models adopted a closure of the 
form (\ref{rpa}) with or without the additional optimization of the potential inside the
hard sphere diameter to enforce the core condition. 

\subsection{HRT in Smooth cut-off and the Self Consistent Ornstein Zernike Approximation}

The Self consistent Ornstein Zernike Approximation (SCOZA) is a well known
liquid state theory which proved successful in the determination of phase diagrams in
fluids characterized by a pair interaction written as the sum of a singular hard sphere part 
plus a Yukawa tail $w(\br)$ \cite{scoza}. Originally, SCOZA has been formulated as a generalization
of the Mean Spherical Approximation \cite{hansen} 
for the direct correlation function outside the core:
\begin{equation}
c(\br)= \lambda\,\phi(\br)
\label{scoza}
\end{equation}
while enforcing the core condition $g(\br)=0$ to define $c(\br)$ inside the hard sphere 
diameter. The parameter $\lambda$ 
is determined by requiring thermodynamic consistency
between the internal energy and the compressibility routes to thermodynamics:
\begin{eqnarray}
\frac{\partial }{\partial\beta} \,\left ( \frac{-\beta {\cal A}_Q}{V}\right )
&=& - \frac{\rho^2}{2} \int d\br \,g(r)\, w(r) \nonumber \\
\frac{\partial^2 }{\partial\rho^2} \,\left ( \frac{-\beta {\cal A}_Q}{V}\right )
&=& \int d\br \,c(r) -\frac{1}{\rho}
\label{scozac}
\end{eqnarray}
The two exact expressions (\ref{scozac}), together with the core condition
and the parametrization (\ref{scoza}) outside the hard sphere diameter, 
give rise to a partial differential
equation for the free energy as a function of $\rho$ and $\beta$.
Remarkably, this equation is exactly reproduced by HRT 
in the smooth cut-off formulation (\ref{smooth0})
by use of a linear turning on of the interaction (\ref{wqsmooth}): 
\begin{equation}
w_t(\br) = t\, w(\br) 
\end{equation}
where $t$ is the switching-on parameter \cite{smooth}. 
Clearly this form does not satisfy the requirements
stated in Section \ref{sechrt} for a selective introduction of density fluctuations: 
while SCOZA just tunes the amplitude of $w(r)$, HRT changes
simultaneously both amplitude and range of the potential by imposing a cut-off $Q$ on 
its long wave-length Fourier components thereby qualifying as a microscopic implementation 
of RG ideas. In fact, SCOZA has been shown to provide a good description of the non 
universal properties in the critical region, as well as non classical critical exponents 
but fails in reproducing the correct scaling picture predicted by RG and HRT \cite{scoza2}. 
An attempt to reconcile HRT and SCOZA has been formulated in Ref. \cite{reinerhoye}, where 
a further adjustable parameter is added in the RPA-like closure of the HRT equation, to enforce 
thermodynamic consistency between the compressibility and the internal energy route.

\section{Critical properties and phase coexistence}

One of the goals which motivated the formulation of HRT was to provide a 
description of the critical properties of realistic fluid models consistent with the 
RG picture. Critical exponents, governing the singularities
of thermodynamic quantities in the critical region, acquire universal 
values quoted in Table \ref{table1} \cite{field}.
At the critical density, the constant volume specific heat diverges as $C_V\div t^{-\alpha}$
as the reduced temperature $t=T/T_c -1$ vanishes; the isothermal compressibility 
as $\kappa \div t^{-\gamma}$ and the correlation length as $\xi\div t^{-\nu}$. 
The shape of the coexistence curve is asymptotically given by $|\rho(T)-\rho_c|\div |t|^{\beta}$,
while the equation of state on the critical isotherm is given by 
$|P-P_c| \div |\rho-\rho_c|^{\delta}$. Finally, precisely at the critical point, 
the pair correlation displays a power law tail of the form $h(r) \div 1/r^{d-2+\eta}$.

In the sharp cut-off formulation, universality 
is indeed a direct consequence of the asymptotic form of the HRT hierarchy
close to the critical point \cite{hrt0}: all the microscopic details of the model 
disappear from the asymptotic equations and the full hierarchy acquires a universal form
equivalent to a previously studied RG differential generator \cite{nicoll}. 
Scaling laws and non classical values for the critical exponents immediately 
follow from this identification. 
In the smooth cut-off formulation, the universality of critical phenomena is less 
apparent because the 
specific form of the chosen cut-off function $\omega_Q(\br)$ (\ref{wqsmooth},\ref{mass}) 
survives also in the asymptotic critical region. Universality of the critical 
properties can then be proved only order by order in the $\epsilon=4-d$ expansion 
and no general proof is available \cite{ap}. Here we will briefly review the 
critical properties of the approximate non perturbative class of closures previously 
introduced both in the sharp and in the smooth cut-off formulation (\ref{rpa}). Moreover we 
will sketch the argument leading to the conclusion that already at the level of 
this simple approximation, HRT always gives a convex free energy, even at 
coexistence. The unphysical divergence of the compressibility at the phase
boundary predicted by the sharp cut-off closure \cite{max1} is circumvented by the 
smooth cut-off formulation which then provides a satisfactory description of 
first order transitions \cite{max2}.

Other, more sophisticated closures have been proposed. As previously noticed, the second
equation of the HRT hierarchy describes the evolution of the two particle
direct correlation function in terms of the three
and four particle correlations (\ref{csmooth},\ref{csharp}). 
By use of the closure (\ref{c3},\ref{c4}) it has been shown
that the critical exponents are correctly reproduced to second order in $\epsilon=4-d$,
including a non vanishing anomalous dimension $\eta$ \cite{hrt0,ap,adv}. 
Unfortunately this quite promising 
non perturbative closure has not been thoroughly investigated in microscopic models, 
probably due to the 
problems posed by the numerical solution of the resulting non linear partial differential equation. 
Recently, a similar (although simpler) approximation has been studied in the 
asymptotic regime, in the framework of
NPRG, leading to encouraging results for the universal quantities \cite{bmw2}. 

\subsection{Sharp cut-off}

In the sharp cut-off HRT, the free energy evolution equation (\ref{sharp})
depends on the two point function ${\cal F}_Q(\bq)$ 
evaluated ``on shell", i.e. on the surface $q=Q$. This means that in the 
late stage of the integration, when $Q$ is small, only the long wave-length 
physics contributes to the free energy flow. Moreover, in the critical region, 
${\cal F}_Q(\bq)$ diverges at small $q$ and then we can disregard the additive 
term in Eq. (\ref{sharp}) and the sub-leading $\log\tilde\phi(\bq)$ contribution.
Finally, the evolution equation acquires a universal form 
\begin{equation}
\frac{1}{V}
\frac{\partial (-\beta {\cal A}_Q)}{\partial Q} = - Q^{d-1}\,\frac{d\,K_d}{2}\,\, \log {\cal F}_Q (Q)
\label{sharp2}
\end{equation}
where $K_d$ is a geometric factor equal to the ratio between the volume of the unit sphere 
in $d$ dimensions
and $(2\pi)^d$. When we insert a closure of the form (\ref{rpa}) at right hand side, recalling
Eq. (\ref{calf}), we need to evaluate the direct correlation function at wave-vector $Q\to 0$.
By exploiting the analyticity of $c_R(\bq)$ and $\phi(\bq)$ around $\bq\sim 0$, we can write
\begin{equation}
{\cal F}_Q(Q) \to \frac{\rho}{1-\rho\,{\cal C}_Q(0) + \rho\,b\,Q^2}
\label{oz2}
\end{equation}
where $b$ determines the curvature of the direct correlation function in this approximation
and attains a finite, positive limit at the critical point.  
Recalling the compressibility sum rule (\ref{komp}) and inserting this asymptotic form 
into Eq. (\ref{sharp2}) we obtain a closed partial differential equation expressing
the free energy flow at long wave-length and in the critical region. The RG structure 
of this equation can be highlighted by a suitable rescaling, as shown in Refs. \cite{hrt0,adv},
which allows the explicit calculation of the universal properties, like 
critical exponents and scaling functions, via the fixed point analysis. The analyticity
in the momentum dependence of the direct correlation function, implied by the 
RPA-like ansatz (\ref{rpa}), immediately gives the vanishing of the anomalous dimension,
or Fisher exponent, $\eta=0$ in any spatial dimension. This result is exact for $d\ge 4$,
when the critical exponents are classical, while for $d<4$ 
such a  simple non perturbative approximation  
to the first HRT equation gives critical properties correct to first order 
in the $\epsilon=4-d$ expansion. However this closure  leads to a qualitatively 
incorrect prediction in $d=2$, when the anomalous dimension is crucial in 
ensuring the existence of a fixed point solution. 
Few critical exponents in $d=3$ are reported in Table \ref{table1} together with
the accepted values obtained from extrapolations of the high temperature 
expansion in the Ising model \cite{field}. 
\begin{table}
\vskip 0.2cm
\begin{tabular}{|c||c|c|c|c|c|c|c|}
\hline
Exponent & $\alpha$   & $\beta$   & $\gamma$ & $\nu$
& $\delta$    & $\eta$ & $U_2=C_+/C_-$ \\
\hline
``Exact"   & 0.110 & 0.327 & 1.237 & 0.630  & 4.789 & 0.036 & 4.76  \\
\hline
HRT-sharp & -0.07 & 0.345 & 1.378 & 0.689 & 5  & 0 & $0$ \\
\hline
HRT-smooth & 0.05 & 0.330 & 1.300 & 0.650 & 5 & 0 &  4.7  \\
\hline
\end{tabular}
\caption{HRT estimates of the critical exponents and compressibility
amplitude ratio in three dimensions for the sharp and the smooth formulations of HRT
compared to the exact values \cite{field} obtained by extrapolation
of high-temperature series expansions. The smooth cut-off procedure adopted here is
detailed in Section \ref{secsmooth}.}
\label{table1}
\end{table}

It has been realized in Ref. \cite{max1} that 
the class of non-perturbative closures to the first HRT equation
previously introduced (\ref{rpa}) gives rise to a convex free energy
in the limit $Q\to 0$: long wave-length fluctuations are responsible
for the flattening of the isotherms in the coexistence region. A similar conclusion 
has been reached in the framework of NPRG in Ref. \cite{wet2}. An analytical 
argument justifying the convexity of the free energy can be provided starting 
from the asymptotic equation (\ref{sharp2}) previously derived. 
Below the critical temperature, mean field approximation gives 
negative compressibility in a finite region around the critical density $\rho_c$, and
mean field approximation acts as initial condition to the HRT flow equation,
which describes the effects of fluctuations in reshaping the free energy. 
Eq. (\ref{oz2}) shows that ${\cal F}_Q(Q)$ grows large at small $Q$. In this regime
it is convenient to introduce the auxiliary quantity
\begin{equation}
u_Q(\rho) = -\log {\cal F}_Q(Q)=
\log \left [b\,Q^2 - 
\frac{\partial^2 }{\partial \rho^2} \left ( \frac{-\beta {\cal A}_Q}{V}\right ) 
\right ] 
\end{equation}
which satisfies the flow equation 
\begin{equation}
e^{u_Q}\,\frac{\partial u_Q}{\partial Q} = 2\,b\,Q - \frac{d\,K_d}{2}\,Q^{d-1}\,
\frac{\partial^2 u_Q}{\partial \rho^2}
\label{equ}
\end{equation}
When ${\cal F}_Q(Q)$ is large, $u_Q$ becomes large and negative 
and the exponentially vanishing term at left hand side of Eq. (\ref{equ}) can be neglected. 
Then Eq. (\ref{equ}) shows that the only possible asymptotic form of $u_Q(\rho)$ is 
\begin{equation}
u_Q(\rho) = \frac{2\,b}{d\,K_d}\,\left [(\rho-\rho_c)^2 - \Delta^2 \right ] \,Q^{2-d}
\end{equation} 
where $\Delta$ is an integration constant. 
However, this solution is consistent with the assumption
$u_Q(\rho)\to -\infty$ only for $\rho_c -\Delta < \rho < \rho_c +\Delta$
(and $d > 2$): below the critical temperature, the compressibility diverges
within a density interval (coexistence region) of width $\Delta$. 
Clearly, this asymptotic solution breaks down on the binodal, i.e. for 
$\rho-\rho_c = \pm \Delta$. As discussed in Ref. \cite{max1}, a study of Eq. (\ref{equ}) near the 
binodal 
shows that just outside the coexistence curve the inverse compressibility behaves as 
\begin{equation}
\frac{1}{V}\,\frac{\partial^2 (-\beta A)}{\partial \rho^2} \propto - \left ( |\rho-\rho_c| - \Delta
\right )^{\frac{4-d}{d-2}}
\end{equation}
for $d <4$ leading to an unphysical divergence of the compressibility at the phase
boundary. Interestingly, above four dimension the correct behavior is restored and
the compressibility attains a finite limit on the coexistence curve. 

\subsection{Smooth cut-off}
\label{secsmooth}

The analysis of the critical properties within the smooth cut-off formulation of HRT is
slightly more subtle than in the sharp cut-off case. Formally, the free
energy evolution equation (\ref{smooth}) depends on the two point function at all 
wavelengths, contrary to the sharp cut-off case, and some preliminarily consideration
on the cut-off dependence of the function $\tilde\phi_Q(\bk)$ is appropriate in order
to simplify the asymptotic flow. According to the
identification (\ref{mass}) the difference $R_Q(\bk)=\tilde\phi(\bk)-\tilde\phi_Q(\bk)$ acts as a
``mass term" in the associated field theory: as such, it should scale as $Q^2$ times a short range
function of the ratio $k/Q$ \cite{wetterich}. In previous applications of HRT to microscopic models
\cite{ap,prl,molphys} the choice 
$R_Q(\bk)=e^{-2s}\,\tilde\phi(k\,e^s)$ was adopted at small values of
$Q$: the momentum cut-off is proportional to $e^{-s}$ with $s:\,0\mapsto \infty$, while 
the cut-off function is identified as the physical two body interaction. In this way an
efficient suppression of long wave-length density fluctuations is achieved, together with
the correct asymptotic scaling. Moreover, this special choice allowed to set the initial 
condition of the HRT flow at $s=0$, when $\tilde\phi_Q(\bk)$ identically vanishes and
the physical properties of the model reduce to those of the known reference system.
Instead, when HRT was applied to the study of the $\phi^4$ scalar field theory \cite{max2} the 
simpler form 
\begin{equation}
R_Q(\bk) =\tilde\phi(\bk)-\tilde\phi_Q(\bk)=b\,(Q^2-k^2)\,\theta(Q-k) 
\label{litim}
\end{equation}
was adopted. 
As usual $b$ defines the long wave-length behaviour of the direct correlation function
${\cal C}_Q(\bk)\sim {\cal C}_Q(0) - b\,k^2$. 
Such a form of the cut-off function $R_Q(\bk)$, 
closely resembles the choice proposed 
in Ref. \cite{litim} and allows to highlight the main features of the
simple RPA closure to HRT in the smooth cut-off formulation, avoiding many of the technical
inconveniences of other options. 
By substituting Eq. (\ref{litim}) into the HRT flow equation (\ref{smooth}) we obtain \cite{max2}:
\begin{equation}
\frac{1}{V} \frac{\partial (-\beta {\cal A}_Q)}{\partial Q} = -b\,K_d\,Q^{d+1}\,\left [ 
-\frac{\partial^2 }{\partial \rho^2} \left ( \frac{-\beta {\cal A}_Q}{V}\right )
+b\,Q^2 \right ]^{-1}
\label{evolphi}
\end{equation}
The initial condition is set at the ultraviolet cut-off of the $\phi^4$ theory, taken as the 
inverse length unit ($Q=1$): 
\begin{equation}
\frac{\beta {\cal A}_1(\rho)}{V} = 
\frac{\beta {\cal A}_1(\rho_c)}{V}+\,t\,(\rho-\rho_c)^2 + u\,(\rho-\rho_c)^4
\end{equation}
Here $t$ is a measure of the mean field reduced temperature 
$t=(T-T_{mf})/T_{mf}$ while $u$ is the self-interaction defining the bare theory. 
It is convenient to rescale the free energy by $K_d$ and the density by 
$\sqrt{K_d/b}$ to get rid of the dimensional constants in the evolution equation (\ref{evolphi}).
A further rescaling of both the density and the free energy by suitable powers of $Q$ confers
a RG structure to this flow equation, whose fixed point corresponds to the critical singularity. 
A standard stability analysis of the fixed point solution allows to evaluate the 
critical exponents, whose values in $d=3$ are shown in Table \ref{table1}. Similarly to
the sharp cut-off case, the parametrization (\ref{rpa}) provides the correct exponents 
to first order in the $\epsilon=4-d$ expansion. 
Being ${\cal C}_Q(\bk)$ analytic in the wave-vector, 
the anomalous dimension vanishes ($\eta=0$) for all $d$ as in the sharp cut-off case.

\begin{figure}
\includegraphics[height=6cm,width=6cm,angle=0]{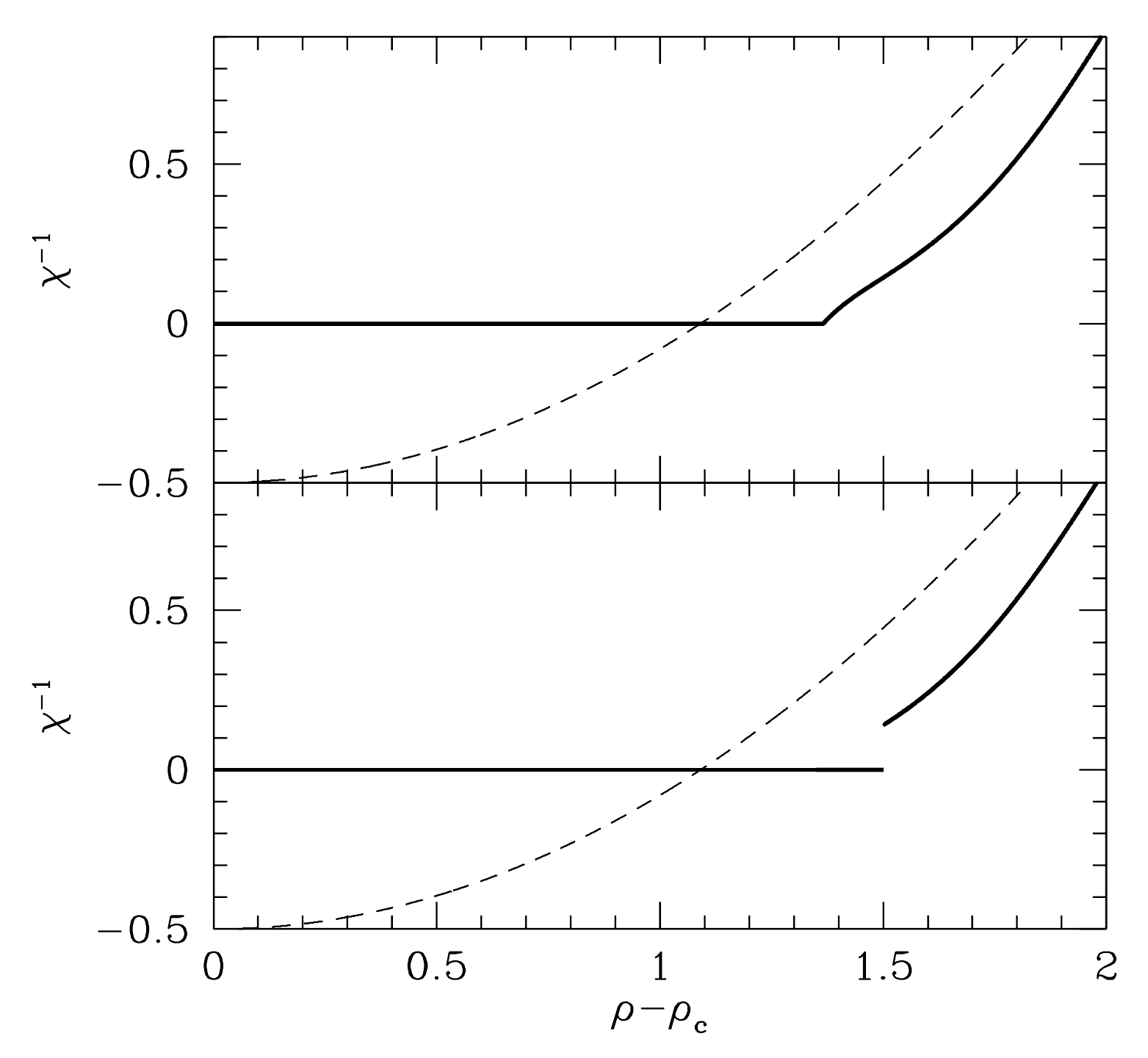}
\caption{Inverse compressibility (\ref{chiq}) of the $\phi^4$ theory in $d=3$ 
below the critical temperature. 
The dashed line corresponds to the initial condition, which coincides with
the mean field approximation, while the solid line is the result at convergence ($Q=0$). 
Data refer to $u=0.035$ and $t=-0.25$. 
Upper (lower) panel shows the results of the sharp (smooth) cut-off formulation of HRT. 
}
\label{figmax}
\end{figure}

In Refs. \cite{catania} it was suggested that the smooth cut-off formulation of RG 
qualitatively improves on the description of the system close to the phase boundary.
This expectation was later confirmed both within HRT \cite{max2} and NPRG \cite{caillol2}.  
Following a procedure similar to that adopted in the sharp cut-off case, we first define
the auxiliary quantity 
\begin{equation}
u_Q(\rho) = \left [ 
-\frac{\partial^2 }{\partial \rho^2} \left ( \frac{-\beta {\cal A}_Q}{V}\right )
+Q^2 \right ]^{-1}
\end{equation}
which, below the critical temperature, 
diverges to $+\infty$ as $Q\to 0$ within the coexistence region. By writing the flow equation
for $u_Q$ and equating the diverging terms to leading order we get the solution
\begin{equation}
u_Q(\rho) = \left [ \Delta^2 - (\rho-\rho_c)^2\right ] \,Q^{-d}
\end{equation}
where $\rho_c$ and $\Delta$ are integration constants. This solution 
is consistent with the assumption $u_Q\to +\infty$ for $|\rho-\rho_c| < \Delta$
showing that $\rho_c$ and $\Delta$ are the critical density and the width of the
coexistence curve respectively. Therefore, also the smooth cut-off formulation of HRT gives
rise to a convex free energy. Moreover, a detailed analysis of the solution in an 
infinitesimal neighbourhood of the phase boundary $\rho\sim \rho_c \pm\Delta$ shows that 
a discontinuity develops for $Q\to 0$ \cite{max2}. This behaviour is clearly illustrated in Fig. 
\ref{figmax} which shows the evolution of the (running) inverse compressibility
\begin{equation}
\chi_Q^{-1}(\rho) = -
\frac{\partial^2 }{\partial \rho^2} \left ( \frac{-\beta {\cal A}_Q}{V}\right )
\label{chiq}
\end{equation}
contrasting the sharp and the smooth cut-off flows. 
As suggested in Ref. \cite{catania}, the presence of a discontinuity at the phase boundary
is a generic feature of any smooth cut-off scheme and not a peculiarity of the chosen form (\ref{litim}). 
The finite value attained by the isothermal compressibility $\chi$ at the phase boundary allows to introduce 
the compressibility amplitude ratio defined as $U_2=C_+/C_-$ where the $\pm$ label refers to 
the case above or below the critical temperature. 
If $\tau$ is the reduced temperature $\tau=(T-T_c)/T_c$,
\begin{equation}
\chi (\rho^*) \sim C_\pm (\pm \tau)^{-\gamma}
\end{equation}
For $\tau>0$ the density is fixed at its critical
value $\rho^*=\rho_c$ while for $\tau <0$ the compressibility is measured along the phase boundary:
$\rho^*=\rho_c+\Delta(\tau)$. 
This amplitude ratio $U_2$ is known to be universal and the smooth cut-off 
HRT estimate agrees well with high temperature extrapolations, as shown in Table \ref{table1}.
The full equation of state in scaling form can be also evaluated as shown in Ref. \cite{molphys}. 

\subsection{Locating the spinodal within HRT}

\begin{figure}
\includegraphics[height=6cm,width=6cm,angle=0]{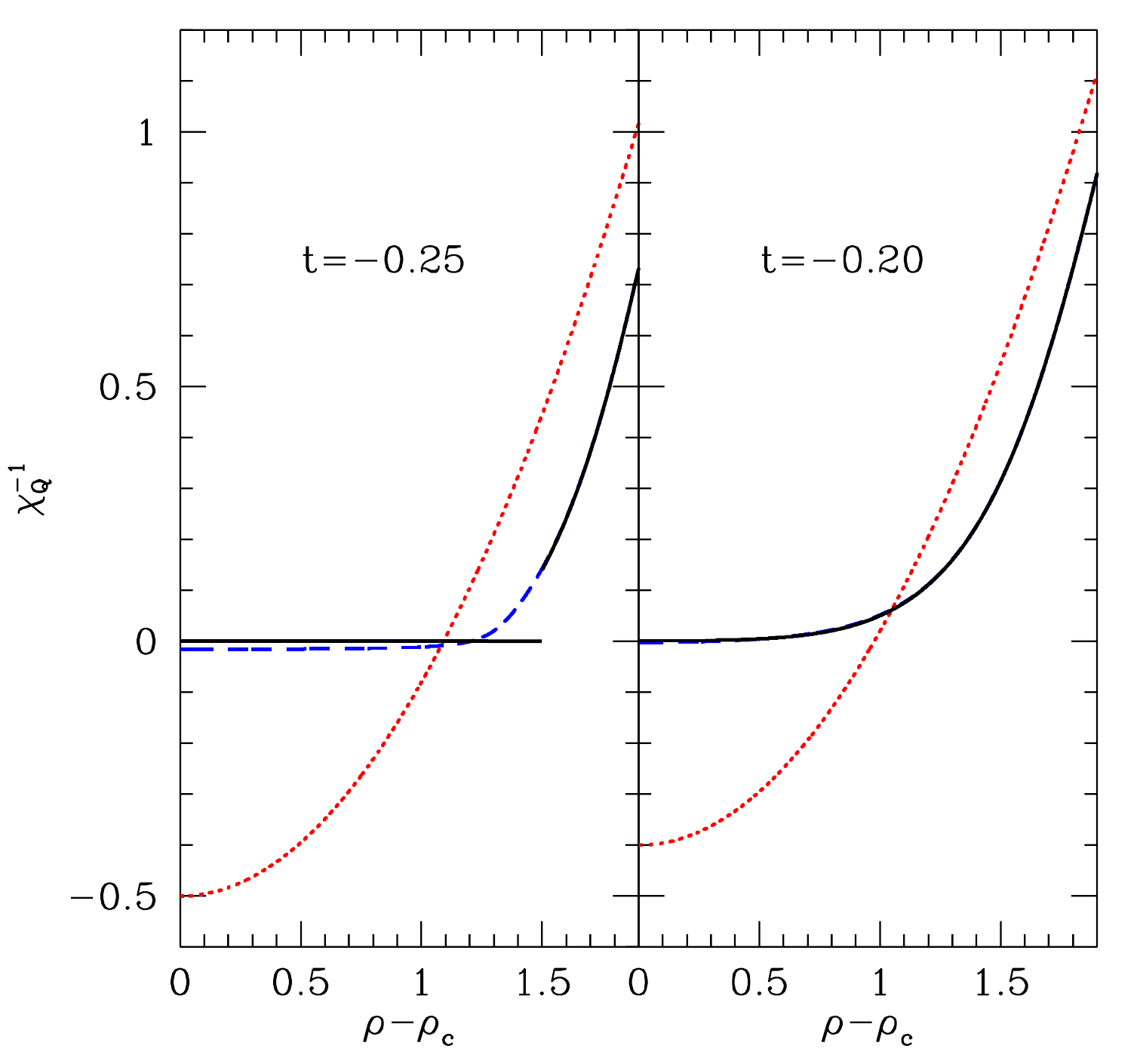}
\caption{Snapshots of the inverse compressibility as a function of density
at three values of the cut-off $Q$ for $u=0.035$. 
Left panel refers to $t=-0.25$ below $t_c\sim -0.203989$,
right panel to $t=-0.20$ above the critical temperature. 
Dotted line (red): $Q=1$ (which coincides with mean field approximation); 
dashed line (blue): $Q=0.135$; full line (black): $Q=0$. 
In the right panel the effects of fluctuations for $Q \lesssim 0.135$
are negligible making the dashed and full lines almost indistinguishable.  
}
\label{figbelow}
\end{figure}

\begin{figure}
\includegraphics[height=6cm,width=6cm,angle=0]{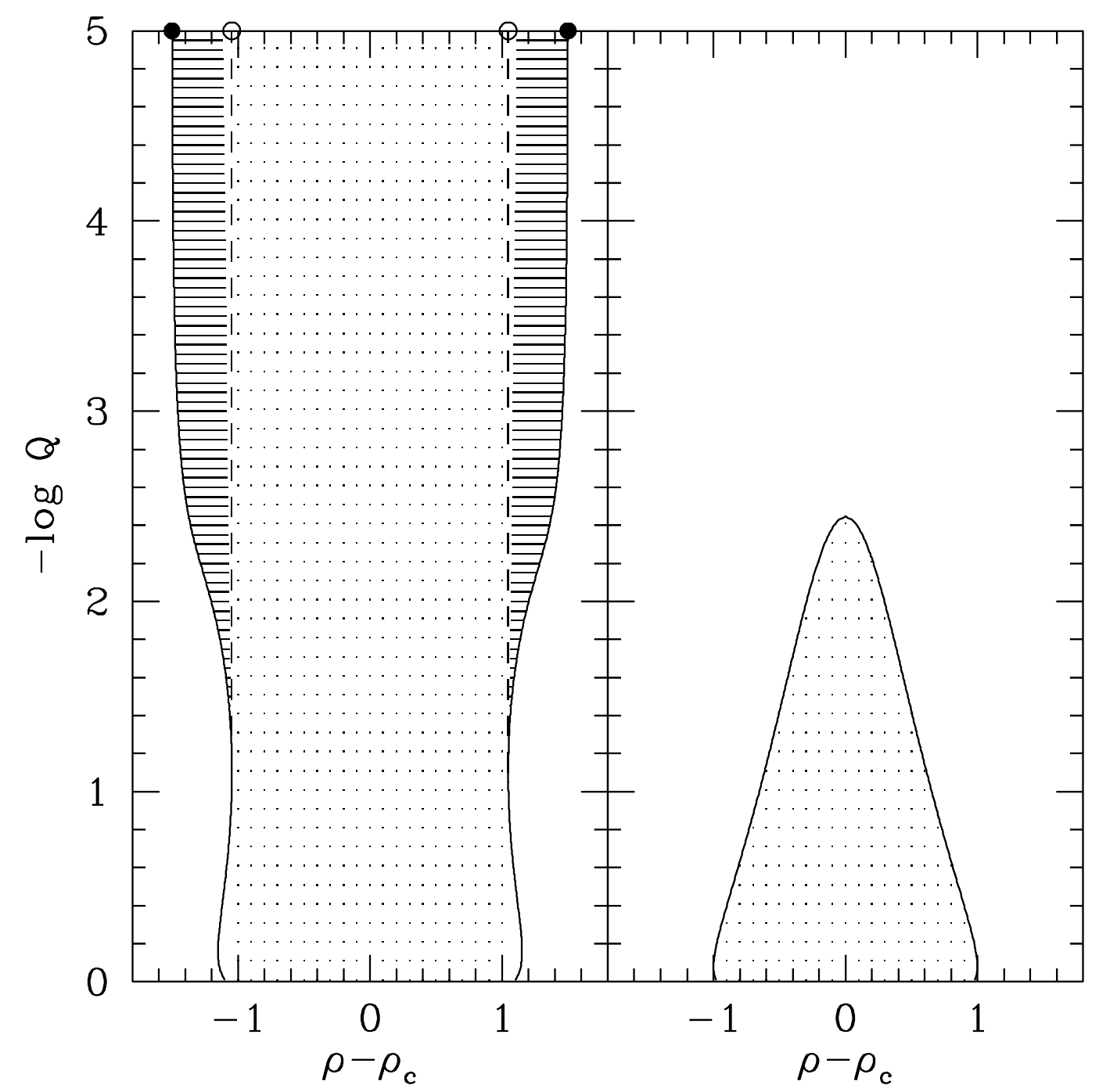}
\caption{Boundaries of the region defined by a negative compressibility as a function
of the cut-off $Q$ for $u=0.035$. At $Q=1$ the standard mean field approximation is recovered,
while all fluctuations are included at the top of the figure.
Left panel refers to $t=-0.25 < t_c$. 
At each $Q$ the shaded area is the metastable region, where the compressibility becomes negative 
due to the inclusion of fluctuations at wavelengths $R_c < \lambda < Q^{-1}$, where $R_c\sim 3$
at this temperature.  
Dots mark the unstable region, characterized by a negative compressibility at all wavelengths
$\lambda < Q^{-1}$.
At convergence (top of the figure) 
full dots mark the coexistence region (located at $|\rho-\rho_c|\sim 1.5$), 
empty dots the ``spinodal" (at $|\rho-\rho_c|\sim 1.05$). 
Right panel, same plot for $t=-0.20 >t_c$. Here the unstable region closes at $Q_c\sim 2.45$, 
and the fully interacting system is in a single phase state. 
}
\label{figrhot}
\end{figure}

The spinodal is a useful way to discriminate between intrinsically unstable and metastable states, 
even if this curve is not rigorously defined within equilibrium statistical mechanics, the
free energy being a convex function of density in the whole phase diagram. 
In Ref. \cite{molphys} it was noticed that 
a special feature of the smooth cut-off HRT flow below the critical temperature 
suggests a way to discriminate between instability and metastability. 
At mean field level, a negative compressibility identifies the unstable 
region lying inside the spinodal curve. However, fluctuations are expected to 
modify the instability boundary which indeed evolves as $Q$ lowers, as 
shown in Fig. \ref{figbelow}: short wave-length fluctuations 
up to a characteristic length-scale $R_c$ slightly 
shrink the instability region, while fluctuations at wave-vector 
$Q \lesssim Q_c=R_c^{-1}$ lead to a widening of the two-phase region,
defined by the asymptotic vanishing of the inverse compressibility.
Therefore the HRT flow allows to identify a density interval, 
deep inside the coexistence region, 
characterized by a negative compressibility at all wave-lengths: the 
unstable region bounded by the spinodal line. 
Instead, close to the phase boundary, but still inside the binodal, a 
region exists where the compressibility remains positive even by including 
fluctuations up to a maximum length-scale $R_c$: the metastable region. 
The curves dividing the regions of positive and negative $\chi_Q^{-1}$
in the $(\rho,Q)$ plane at different temperatures are shown in Fig. \ref{figrhot}, together with
the extent of the metastable region obtained via the criterion previously outlined.
The cut-off value $Q_c$ where the width of the unstable region has a minimum,
vanishes when the critical point is approached. 
This corresponds to the divergence of a characteristic
length-scale $R_c \div 1/Q_c$, which may be identified as the critical droplet radius on the
spinodal line, according to the standard droplet model picture of nucleation \cite{drop}.
This divergence follows a power law consistent with the scaling
$R_c\div \xi \div |\tau|^{-\nu}$ with $\nu=\gamma/2$ (recall that within the 
present approximation the critical exponent $\eta$ vanishes).

\section{Applications to microscopic models}

RPA-like closures of the type discussed in Section \ref{appro} have been applied to several 
microscopic models both in the continuum and on the lattice within the sharp cut-off formulation
of HRT. The earliest studies were 
focused on simple liquids, like noble gases or Lennard-Jones fluids \cite{meroni}. 
The RPA closure (\ref{rpa}) was later suitably
generalized to include the correct short range behavior of 
correlations characterizing the liquid phase,
the so called core condition, which forces the vanishing of the radial distribution 
function inside the hard sphere radius. Extensions of this procedure 
to deal with soft core potentials, were also devised \cite{adv}.  
The Ising model on a cubic lattice was investigated as the inescapable test case of any theory
which qualifies for the description of phase transitions \cite{ising}. More recent applications, together
with the first implementations of the smooth cut-off formulation of HRT and an extension to binary fluids
are briefly reviewed in the following Sections.

\subsection{One component fluids}

The very first physical system studied by HRT (in the sharp cut-off formulation)
was a microscopic model for simple fluids \cite{meroni}, which allowed for a detailed 
comparison with experiments in rare gases both for the thermodynamics and 
for structural data \cite{barocchi}. Later, HRT was also applied to 
complex fluids, a model for $C_{60}$ \cite{tau}, and more
recently to systems of interest in soft matter physics. 
An HRT code in smooth cut-off formulation has been also developed and applied to
the study of a Hard Core Yukawa fluid for different values of the inverse range $z$ \cite{smooth}
and to the Screened Restricted Primitive model of electrolytes \cite{rpm}. 
Here we briefly review three physical systems where the HRT in sharp cut-off formulation
has been successfully applied: a colloidal suspension with added polymers, 
a system of colloidal particles 
with competitive interactions and a model of star polymers in solution. 
The aim of this presentation is to show the 
flexibility of the HRT formalism in dealing with markedly different classes of systems.
The set up of a numerical code for the solution of the HRT non-linear partial differential equation
presents some technical problem due to the occurrence of singularities below the 
critical temperature, related to the divergence 
of the isothermal compressibility in the whole coexistence region. 
These problems, already discussed in the literature \cite{tau,reiner1,reiner2}, 
are also briefly illustrated in the Appendix. 

\begin{figure}
\includegraphics[height=6cm,width=6cm,angle=0]{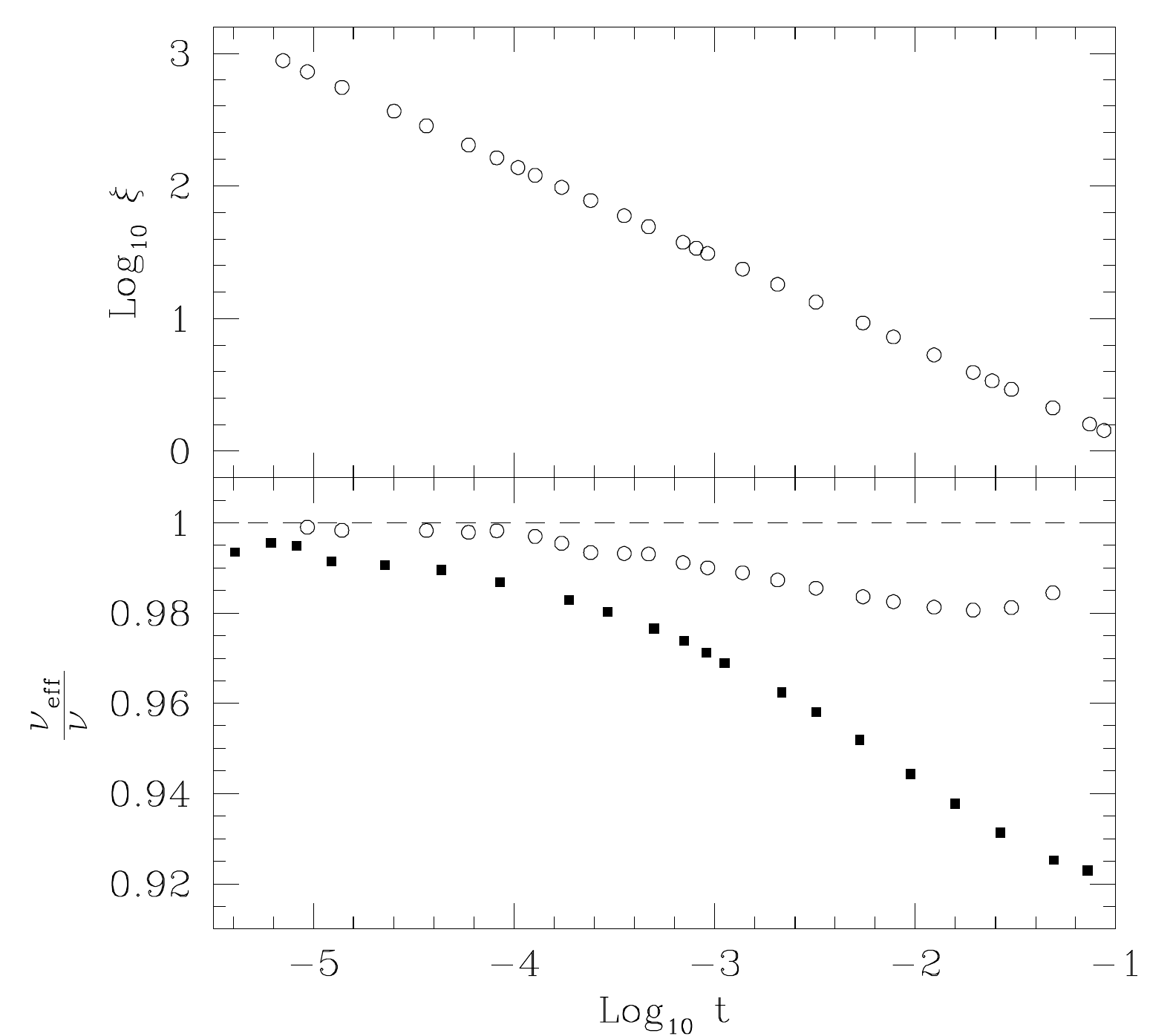}
\caption{HRT results in sharp cut-off formulation for the AO model (\ref{ao}). 
Upper panel: true correlation length $\xi$ as a function of the reduced 
temperature $t=\frac{T}{T_c}-1$ in a logarithmic plot. Its slope defines the critical exponent $\nu$. 
Lower panel: effective critical exponent $\nu_{eff}$, normalized to its asymptotic
HRT value $\nu=0.69$, for the AO model (open circles) and for a Hard Core-Yukawa 
fluid (full squares).  
}
\label{figao}
\end{figure}

Depletion interactions are known to be ubiquitous in soft matter physics. The most celebrated 
system where the depletion mechanism is able to induce particle aggregation is a suspension
of colloids (modeled as hard spheres) where some non-ionic depletion agent, usually a polymer, is added.
The depletant is treated as an ideal gas, while an excluded volume interaction acts between 
polymers and colloids. The effective colloid-colloid pair interaction turns out to be 
attractive and its analytical form can be calculated exactly in this model \cite{barrat}: 
\begin{equation}
w(r)=-kT\, \eta_p\,\frac{(1+q)^3}{q^3}\left [ 1-\frac{3\,r}{2\,(1+q)} + 
\frac{r^3}{2\,(1+q)^3} \right ]
\qquad \text{for} \quad 1 < r < 1 + q
\label{ao}
\end{equation}
Distances are measured in units of the colloid diameter 
$\sigma$ and $q=2\,R_g/\sigma$, where $R_g$ is the gyration radius of 
the polymer. This effective interaction 
is proportional to the (reservoir) polymer packing fraction $\eta_p$ and, 
being of entropic origin, 
to the temperature. HRT in sharp cut-off formulation 
has been tested against numerical simulations for this model 
\cite{vink} using the hard sphere gas as reference system and 
the Asakura-Oosawa (AO) potential (\ref{ao})
as attractive tail. The very short ranged potentials which occur quite often in soft matter physics
cannot be accurately represented by use of the RPA-like non-perturbative closure (\ref{rpa}) 
adopted in most of the implementations of HRT. Therefore, in the study of the Asakura-Oosawa model
moderately large size ratios, $q\ge 0.4$, have been considered. For these systems many body 
interactions, which appear for $q > 0.154$,  become relevant and affect the location of the 
coexistence curve, as shown in the numerical study performed in Ref. \cite{vink}. 
However, the comparison between simulations and HRT results for the AO model shows a good 
agreement and allows to quantify the extent of the critical region, also 
characterizing the crossover 
between a mean-field classical pre-critical region and the asymptotic regime. 
A nice power-law divergence of the correlation length in an extended range of reduced temperatures
is shown in the upper panel of Fig. \ref{figao}, while in the lower panel 
the temperature dependence of the effective
critical exponent $\nu$ is compared to the case of a standard 
Hard Core-Yukawa fluid with $z=1.8$,
which is known to mimic the properties of simple fluids. Interestingly, the AO model displays
a power-law behaviour in a significatively larger domain of reduced temperatures \cite{loverso2}. 

\begin{figure}
\includegraphics[height=6cm,width=6cm,angle=0]{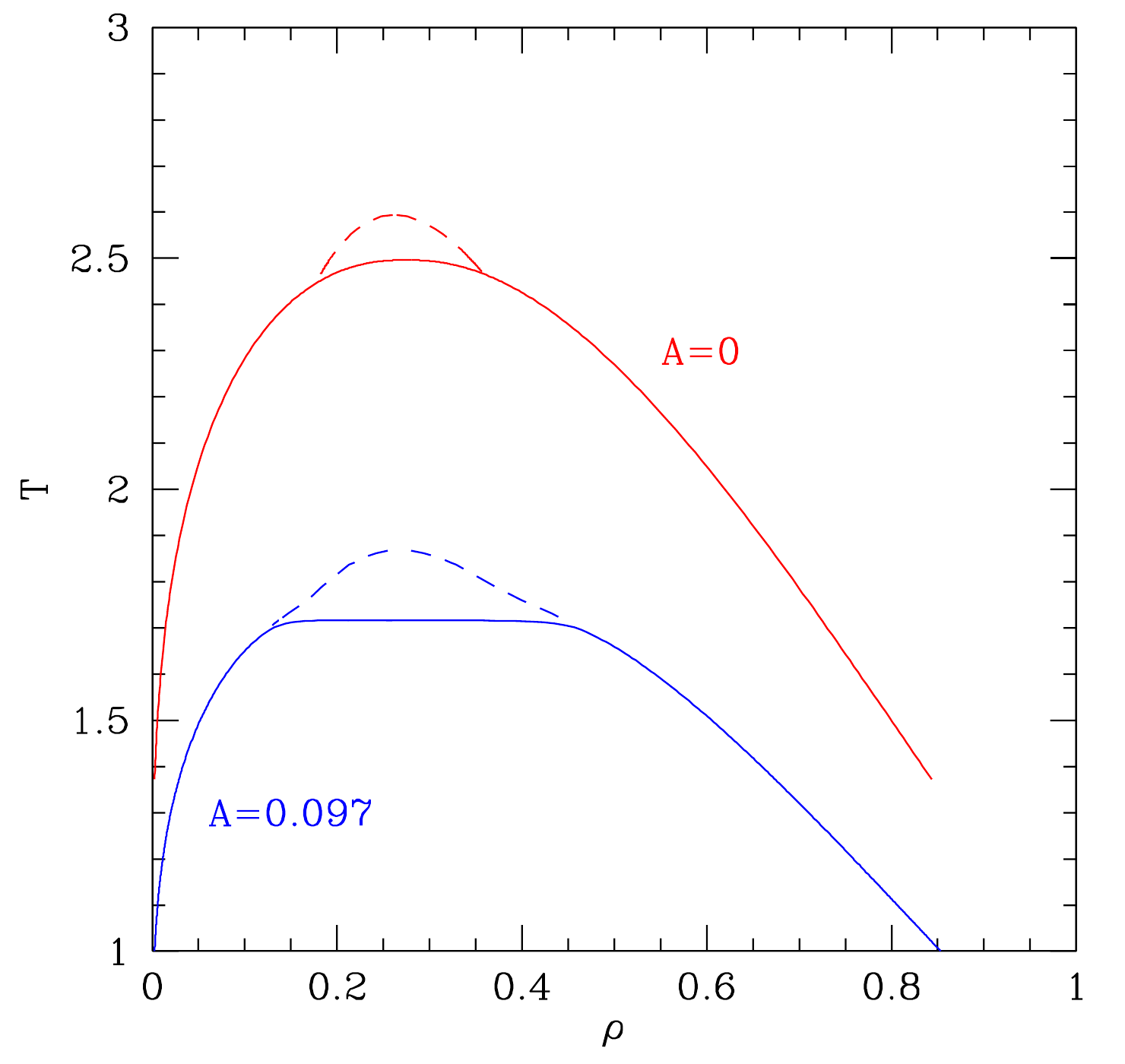}
\caption{
Coexistence regions of the HCTYM, obtained by SCOZA, for two choices of the amplitudes of
the repulsive tail: $A=0$ (red line) and $A=0.097$ (blue line). The dashed lines limit the regions
of enhanced compressibility, defined as the thermodynamic states where the dimensionless isothermal 
compressibility $S(0)$ is lager than $10$. 
}
\label{figcina1}
\end{figure}

\begin{figure}
\includegraphics[height=6cm,width=6cm,angle=0]{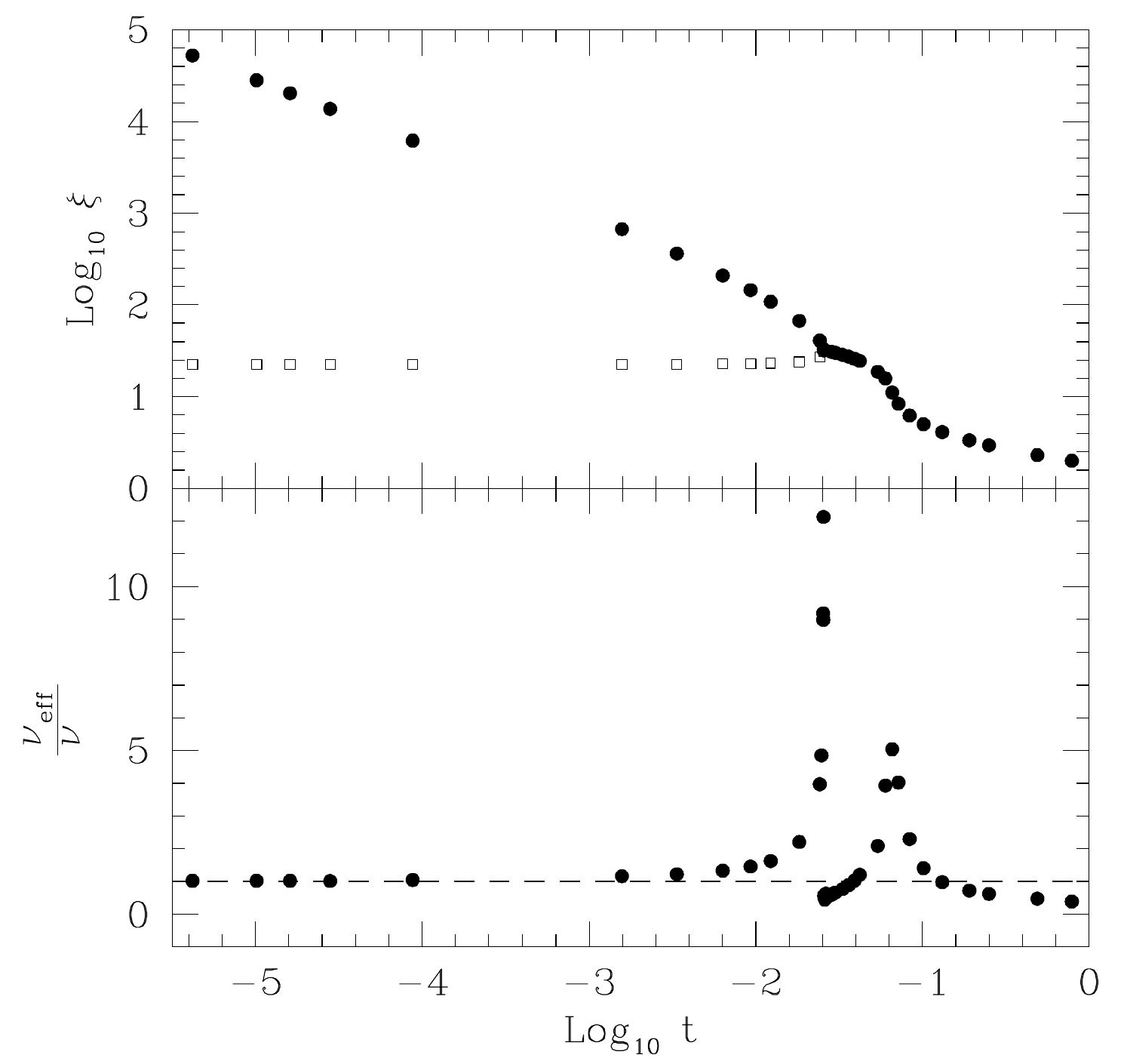}
\caption{
HRT results in sharp cut-off formulation for the HCTY model \cite{cina} at $A=0.097$.
Upper panel: true correlation length $\xi$ as a function of the reduced 
temperature $t=\frac{T}{T_c}-1$ in a logarithmic plot. The diverging correlation 
is shown by full circles. Open squares represent the ``intra-cluster" correlation length. 
Lower panel: effective critical exponent $\nu_{eff}$, normalized to its 
HRT value $\nu=0.69$, for the asymptotically diverging correlation. 
}
\label{figcina2}
\end{figure}

Charged colloids repel each other via a screened Yukawa potential which, 
together with an attractive short ranged attraction, due either to dispersion forces or
to the presence of a depletion agent in solution, give rise to an effective, tunable, 
potential characterized by interparticle attraction at short distances 
and a repulsive tail at longer range. Few experimental results \cite{expcin}, together with
a theoretical analysis based on simulations \cite{imperio}, suggest the possible occurrence
of modulated phases of different morphologies. A class of potentials with these features was
investigated via the sharp cut-off HRT \cite{cina} in the homogeneous fluid phase, where 
the incipient cluster formation may induce peculiar thermodynamic and structural anomalies. 
The pair interaction of the model is of the form named Hard Core Two Yukawa Fluid (HCTYF): 
\begin{equation}
v(r)=
\begin{cases}
+\infty & \text{for} \quad r<1 \\
-\epsilon\,\frac{e^{-z_1(r-1)}}{r} + A\,\epsilon \,\frac{e^{-z_2(r-1)}}{r}
& \text{for} \quad r > 1
\end{cases}
\label{cinapot}
\end{equation}
As usual, distances are measured in units of the hard core diameter. 
The reference system is defined by the hard sphere part plus the repulsive Yukawa tail,
while $w(r)$ is identified with the attractive contribution. The investigation in Ref. \cite{cina}
has been carried out for the choice $z_1=1$ and $z_2=0.5$. Temperatures are measured in units of
the energy scale $\epsilon$ while the parameter $A > 0 $ is varied. 
Within HRT the fluid phase is stable against microphase formation for $A\lesssim 0.097$ but 
even such a small repulsive contribution to the interparticle interaction gives rise to 
remarkable changes to the phase behavior of the fluid. When the parameter $A$ approaches the
stability limit, the coexistence curve considerably flattens near the critical point
and a large region of enhanced compressibility opens above the 
critical temperature. The same picture is confirmed by use of SCOZA, which provides a 
phase diagram topology in quantitative agreement with HRT. 
In Fig. \ref{figcina1} SCOZA results are shown to display such a remarkable flattening of the 
binodal. 
The physical region for such a peculiar behavior
may be identified as the presence of large clusters of strongly correlated particles, which
anticipate the formation of modulated phases. This can be seen in Fig. \ref{figcina2}
where, in the upper panel, the presence of two correlation lengths below a threshold temperature
is illustrated. While a single correlation length does diverge at the critical point, which 
belongs to the Ising universality class, the second, ``intra-cluster", correlation length   
remains remarkably large in the whole critical region. The presence of two correlation lengths 
reflects in the occurrence of anomalies of the effective critical exponent 
on approaching the critical point, as shown in the lower panel of Fig. \ref{figcina2} for
$A=0.097$.
A sharp growth of both correlations above the critical temperature gives rise to strong peaks 
and sharp jumps in
$\nu_{eff}$, while the true asymptotic behavior of the correlation lengths can be seen only 
very close to the critical temperature.  

\begin{figure}
\includegraphics[height=6cm,width=6cm,angle=0]{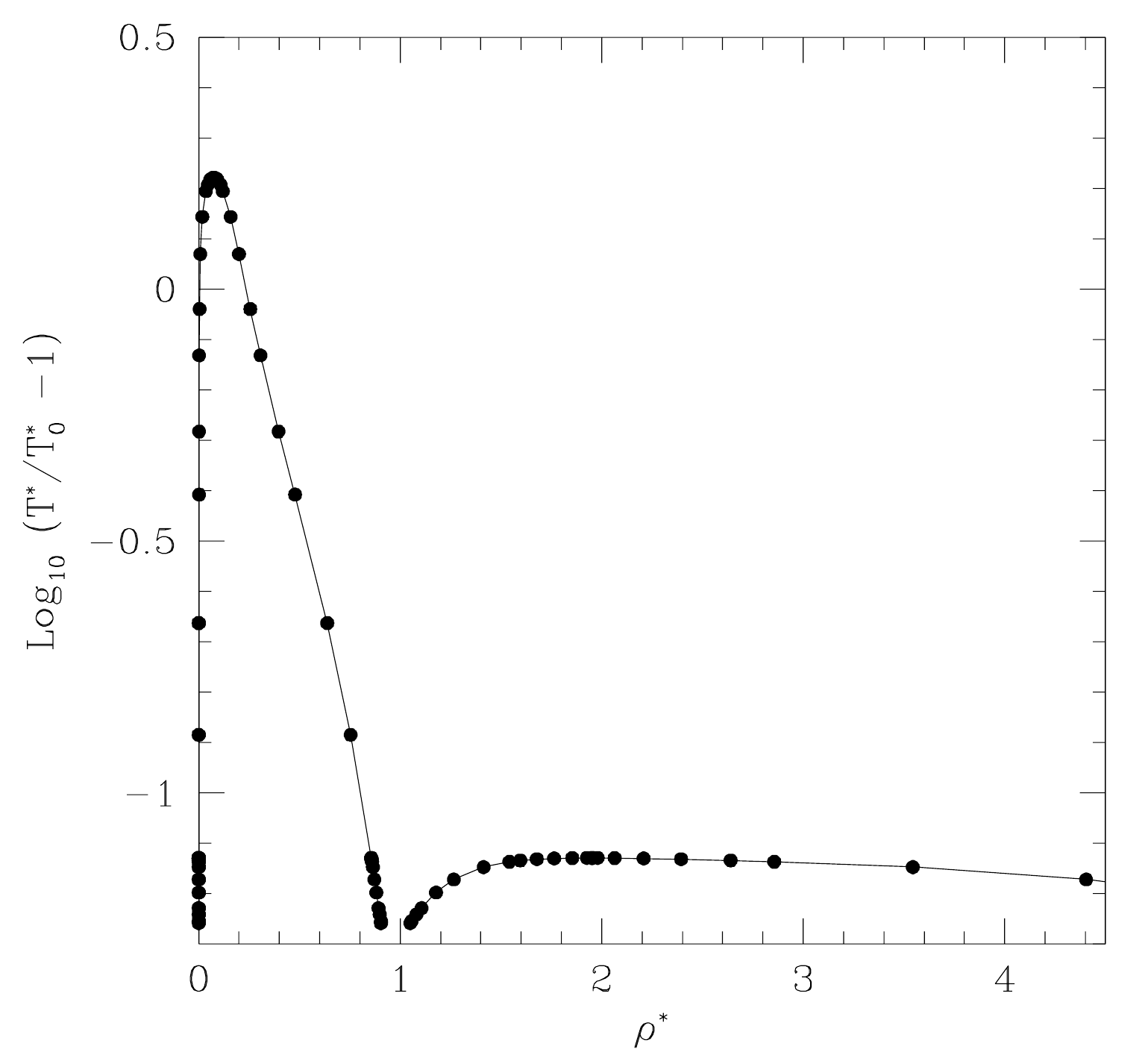}
\caption{HRT coexistence curve for the model of star polymers in 
solution discussed in the text \cite{loverso}. A non-linear temperature scale
has been used to enhance the visibility of the binodal structure (the temperature scale 
has been set for convenience to $T^*_0=0.23$). 
Note the presence of two critical points: one 
at extremely low concentration  $(\rho_{c1}^* = 0.074,T^*_{c1}=0.6123)$
and the other at liquid densities $(\rho_{c2}^* = 1.952,T^*_{c2}=0.2471)$.
The triple point is located at $(\rho_{t}^* = 0.975,T^*_{t}=0.2426)$.
}
\label{figstar}
\end{figure}

Star polymers in a good solvent display an effective interaction
characterized by an ultra-soft repulsive core of entropic origin, 
which can be conveniently parametrized as: 
\begin{equation}
v_R(r) = 
\begin{cases}
kT \, \frac{5}{18} f^{3/2}\left [ 
-\ln r +\left (1+\frac{\sqrt{f}}{2}\right )^{-1}\right ] & \text{for} \quad r < 1 \\
kT \, \frac{5}{18} f^{3/2} \left (1+\frac{\sqrt{f}}{2}\right )^{-1} \, 
\frac{e^{\frac{1}{2}\sqrt{f}(1-r)}}{r} & \text{for} \quad r > 1
\end{cases}
\label{soft}
\end{equation}
where distances are measured in units of the corona diameter of the star $\sigma$ and the 
functionality $f$ of a star polymer coincides with the number of its arms. This 
expression accurately reproduces Molecular Dynamics simulations for $f\gtrsim 10$ \cite{likos}.
Here we will discuss the case $f=32$, displaying a stable fluid phase at
all concentrations $\rho^*$ (in units of $\sigma^{-3}$). 
HRT has been used to study the consequences of an additional attractive interaction
between star polymers, on top of the entropic repulsive contribution (\ref{soft}): 
such an attraction may be induced by different
physical mechanisms, like the presence of residual dispersion forces or the effects of 
a depletant added in the solution. Recent simulations \cite{camargo} confirmed that the attractive 
interaction between star polymers due to added depletant can be conveniently 
parametrized by a Fermi function: 
\begin{equation}
w(r) = -C\,\left [ e^{\frac{r-A}{B}}+1\right ]^{-1}
\end{equation}
This model was previously studied by means of the sharp cut-off formulation of HRT \cite{loverso}
for $f=32$ and the choice of the parameters $A=2.1$, $B=0.35$, while $C$ 
played the role of a temperature scale which defines the 
dimensionless temperature $T^* = kT/C$. For this model the reference system, defined by
the potential (\ref{soft}), is highly non trivial and its thermodynamic and structural 
properties, which serve as an input to the HRT formalism, 
were obtained by use of accurate integral equations of Liquid State Theory.
In the investigation carried out in Ref. \cite{loverso} the Modified Hypernetted Chain 
equation \cite{hansen} was adopted.  
The resulting phase diagram is shown in Fig. \ref{figstar}
and displays two critical points separated by a triple point. Both the 
extremely low critical concentration $\rho_{c1}^*$ and the very presence of the high density 
phase transition are consequences of the ultra-soft nature of the repulsive effective interaction. 

\subsection{Lattice models}

Applying HRT to lattice models requires a few 
preliminary modifications of the evolution equations due to the different nature of the
order parameter and to the geometry of the Brillouin Zone (BZ). 
Let us consider an interacting spin model on a lattice, defined by the hamiltonian
\begin{equation}
H = - \sum_{\bR,\bR^\prime} J(\bR-\bR^\prime)\,\bS_\bR \cdot \bS_{\bR^\prime} - h\sum_R S^z_\bR
\label{ising}
\end{equation}
where $\bS_\bR$ are $n$-component unit vectors defined on the sites $\bR$ of a lattice.
The spin coupling $J(\bR)$ and the external magnetic field $h$ in the $z$ direction are 
the parameters defining the model. We will consider the case of un-frustrated
ferromagnetic systems ($J(\bR) > 0$), where the order parameter of the transition
can be identified as the magnetization per site and can be written 
as a thermodynamic derivative of the partition function $Z(h)$:
\begin{equation}
m = \langle S^z_\bR \rangle = \frac{1}{N}\,\frac{\partial \log Z(h)}{\partial \beta h}
\end{equation}
In anti-ferromagnetic models ($J(\bR) < 0$) 
it would be appropriate to introduce a symmetry breaking 
staggered external field instead of $h$. 
According to the basic principles of HRT, 
we define a sequence of interactions $J_Q(\bR)$, labeled by an 
infra-red cut-off $Q$ which suppresses 
the long wavelength Fourier components, thereby inhibiting the order parameter 
fluctuations on those scales. As usual, it is convenient to introduce the cut-off dependent 
free energy $A_Q(m)$ by performing a Legendre transform on $\log Z(h)$. Then we define 
a modified free energy ${\cal A}_Q$ by including the full mean field contribution (\ref{leg}):
\begin{equation}
{\cal A}_Q(m) = A_Q(m) 
- \frac{m^2}{2} \sum_{\bR,\bR^\prime}
\left [ J(\bR-\bR^\prime) - J_Q(\bR-\bR^\prime)\right ] 
\label{leg2}
\end{equation}
The free energy evolution equation at fixed temperature and magnetization $m$ 
(along the $z$ axis) can be immediately obtained by first order perturbation theory:
\begin{equation}
\frac{1}{N}
\frac{\partial {\cal A}_Q }{\partial Q} = -\frac{1}{2}\,\sum_{\bR}\,
F_Q(\bR) \,\frac{\partial J_Q(\bR)}{\partial Q} 
\label{lsmooth0}
\end{equation}
where $F_Q(\bR)$ is the connected spin-spin correlation function  
\begin{equation}
F_Q(\bR-\bR^\prime) = \langle \bS_\bR \cdot \bS_{\bR^\prime}\rangle_Q - 
m^2
\label{iso}
\end{equation}
and the canonical average $\langle\cdots\rangle_Q$ is taken at fixed 
magnetization $m$ and interaction $J_Q(\bR)$.
Eq. (\ref{iso}) 
can be written in terms of the longitudinal $F_Q^\parallel(\bR)$ and transverse 
$F_Q^\perp(\bR)$ correlations 
\begin{eqnarray}
F^\parallel_Q(\bR-\bR^\prime) &=& 
\langle S^z_\bR S^z_{\bR^\prime}\rangle_Q - 
m^2
\\
F^\perp_Q(\bR-\bR^\prime) &=& 
\langle S^x_\bR S^x_{\bR^\prime}\rangle_Q 
\end{eqnarray}
with the result
\begin{eqnarray}
\frac{1}{N}
\frac{\partial {\cal A}_Q }{\partial Q} &=& -\frac{1}{2}\,\sum_{\bR}\,
\left [ F^\parallel_Q(\bR) + (n-1)\,F^\perp_Q(\bR)\right ]\,\frac{\partial J_Q(\bR)}{\partial Q}  \\
&=& -\frac{1}{2}\,\int_{BZ} \frac{{\rm d}\bk}{(2\pi)^d} \, 
\left [ F^\parallel_Q(\bk) + (n-1)\,F^\perp_Q(\bk)\right ]\,\frac{\partial \tilde J_Q(\bk)}{\partial Q}
\label{lsmooth1}
\end{eqnarray}
where the integral is restricted to the appropriate Brillouin Zone of the chosen 
$d$-dimensional lattice. 

Following the same steps detailed in Section \ref{sec:exact} we first define new correlation functions 
by including in $F^{\alpha\alpha}_Q(\bk)$ the mean field contribution due to the remaining part of the 
interaction $\tilde J(\bk)-\tilde J_Q(\bk)$: 
\begin{equation}
{\cal F}^{\alpha\alpha}_Q(\bk) = \frac{F^{\alpha\alpha}_Q(\bk)}{1-\beta\,F^{\alpha\alpha}_Q(\bk)
\left [ \tilde J(\bk)-\tilde J_Q(\bk)\right]}
\end{equation}
Then we recall the thermodynamic relations expressing the 
longitudinal and transverse susceptibilities in terms of
the corresponding correlation functions at vanishing momentum: 
\begin{eqnarray}
\left [ {\cal F}^{\parallel}_Q(0)\right]^{-1} &=& 
\left [\chi_Q^\parallel\right ]^{-1}=
\beta\,\frac{\partial^2}{\partial m^2} \left ( \frac{{\cal A}_Q(m)}{N}\right )
\nonumber \\
\left [{\cal F}^{\perp}_Q(0)\right]^{-1} &=& 
\left [ \chi_Q^\perp \right ]^{-1} =
\frac{\beta}{m}\,\frac{\partial}{\partial m} \left ( \frac{{\cal A}_Q(m)}{N}\right )
\label{sumrule}
\end{eqnarray}
And finally, we parametrize these correlation functions in a simple RPA-like form, analogous to 
Eq. (\ref{rpa}) 
\begin{equation}
{\cal F}^{\alpha\alpha}_Q(\bk) = \frac{kT}{\lambda^\alpha - \tilde J(\bk)}
\label{lrpa}
\end{equation}
which defines a non-perturbative closure to the evolution equation, provided the parameters 
$\lambda^{\parallel}$ and $\lambda^{\perp}$ are determined by the exact sum rules (\ref{sumrule}). 
Analogously to the previously examined continuum case, it is convenient to define the 
cut-off dependent interaction $\tilde J_Q(\bk)$ by an expression of the form 
of Eq. (\ref{wqsmooth})
where the $\theta$-function selects a surface where 
the interaction $\tilde J(\bk)$ is constant.
In the sharp cut-off limit, a possible choice is: 
\begin{equation}
\tilde J_Q(\bk) = 
\begin{cases}
\tilde J(\bk) & \text{for } 
\tilde J(\bk)  < \tilde J(0)\,(1-Q) \\
0 & \text {elsewhere}
\end{cases}
\label{sigma}
\end{equation}
For  $Q=1-\text{Min }\,\,\tilde J(\bk)/\tilde J(0)$ the system reduces to non-interacting spins.
By lowering $Q$, more and more Fourier components of $J(\bk)$ are included into the
free energy until, in the $Q\to 0$ limit, the full interaction is recovered. 
With such a choice, 
the HRT flow equation for the free energy (\ref{lsmooth0}) in sharp cut-off acquires the form:
\begin{equation}
\frac{1}{N}
\frac{\partial {\cal A}_Q }{\partial Q} = \frac{kT}{2}\,D(Q)
\left [ 
\log\frac{[\chi_Q^\parallel ]^{-1}+\beta\tilde J(0)}
{[\chi_Q^\parallel]^{-1}+Q\,\beta\tilde J(0)}
+(n-1) \log\frac{[\chi_Q^\perp ]^{-1}+\beta\tilde J(0)}
{[\chi_Q^\perp]^{-1}+Q\,\beta\tilde J(0)}
\right ]
\label{lsharp}
\end{equation}
where $D(Q)$ is the ``density of states" defined by
\begin{equation}
D(Q) = \int_{BZ} \,\frac{{\rm d}\bk}{(2\pi)^d} \,\delta\left (
\textstyle{\frac{\tilde J(\bk)}{\tilde J(0)}}
-1+Q\right )
\end{equation}
The critical properties provided by this equation are exact to
first order both in the $\epsilon=4-d$ and in the $1/n$ expansion. The HRT results for the 
correlation length exponent $\nu$ in three dimensions, obtained by 
numerically solving the fixed point equation associated to Eq. (\ref{lsharp}), 
are shown in Table \ref{ltable} for few values of $n$. 
In this approximation the two point correlations are analytic functions of $\bk^2$ and then the
anomalous dimension exponent $\eta$ vanishes, while the theoretical estimates suggest 
$\eta\sim 0.03-0.04$ for all the values of $n$ considered here. The other HRT exponents
can be obtained by use of the scaling and hyperscaling relations which are rigorously verified
within this approximation.
Notice that in the sharp cut-off HRT formalism, the critical exponents 
directly follow from the single assumption of 
analyticity of $\left [{\cal F}_Q(\bk)\right ]^{-1}$ on $\bk^2$ at all $Q$'s \cite{adv}: any 
approximation retaining this property is bound to the values shown in Table \ref{ltable}
which constantly overestimate by about $10\%$ the accepted values. 
HRT is shown to provide the best results for $n=0$, where the $n$-vector model 
describes the self avoiding walk problem, while their quality 
somehow worsens at larger $n$, although they reproduce the correct limit $\nu=1$ as $n\to\infty$. 
\begin{table}
\vskip 0.2cm
\begin{tabular}{|c||c|c|c|c|c|c|}
\hline
Exponent & $n=0$  & $n=1$   & $n=2$  & $n=3$  \\
\hline
HRT-sharp & $0.606$ & $0.689$ & $0.768$   & $0.826$\\
\hline
``Exact" & $0.588$ & $0.630$ & $0.67$  & $0.71$   \\
\hline
\end{tabular}
\caption{The correlation length critical exponent $\nu$ predicted by Eq. (\ref{lsharp}) 
on the basis of a fixed point analysis of the HRT equation in sharp cut-off formulation. 
The ``exact" values are theoretical estimates \cite{field} obtained by combining extrapolations
of high-temperature series expansions and field theoretical resummations. The limiting value
$\nu=1$ for $n \to\infty$ is correctly predicted by HRT. 
}
\label{ltable}
\end{table}

The Ising model on a cubic lattice was the first lattice system investigated by 
sharp cut-off HRT \cite{ising} by use of 
a more sophisticated closure, belonging to the class previously discussed,  
which consistently preserves the spin normalization constraint $\bS_\bR^2 =1$.
A thorough analysis of lattice models via the smooth cut-off
version of HRT has not been attempted yet, although a very similar procedure, put forward
in the framework of non perturbative Renormalization Group,
has been recently developed \cite{dupuis}.
The results shown in \cite{ising} provided a very accurate estimate of the critical temperature
of the Ising model
and compared favourably with the coexistence curve predicted by high temperature expansion.  
A numerical study of the same HRT equation aimed at determining the
effective parameters of a scalar field theory able to reproduce the correct pre-critical
behaviour of the Ising model was performed in Ref. \cite{brognara1}. 
Interestingly, HRT showed that a $\phi^4$ representation of the long wave-length 
spin fluctuations was inadequate for a faithful representation of the non-universal
properties of the Ising model, while the inclusion of a $\phi^6$ term allowed to 
considerably improve the description of the critical region of the model. 

\begin{figure}
\includegraphics[height=7cm,width=7cm,angle=0]{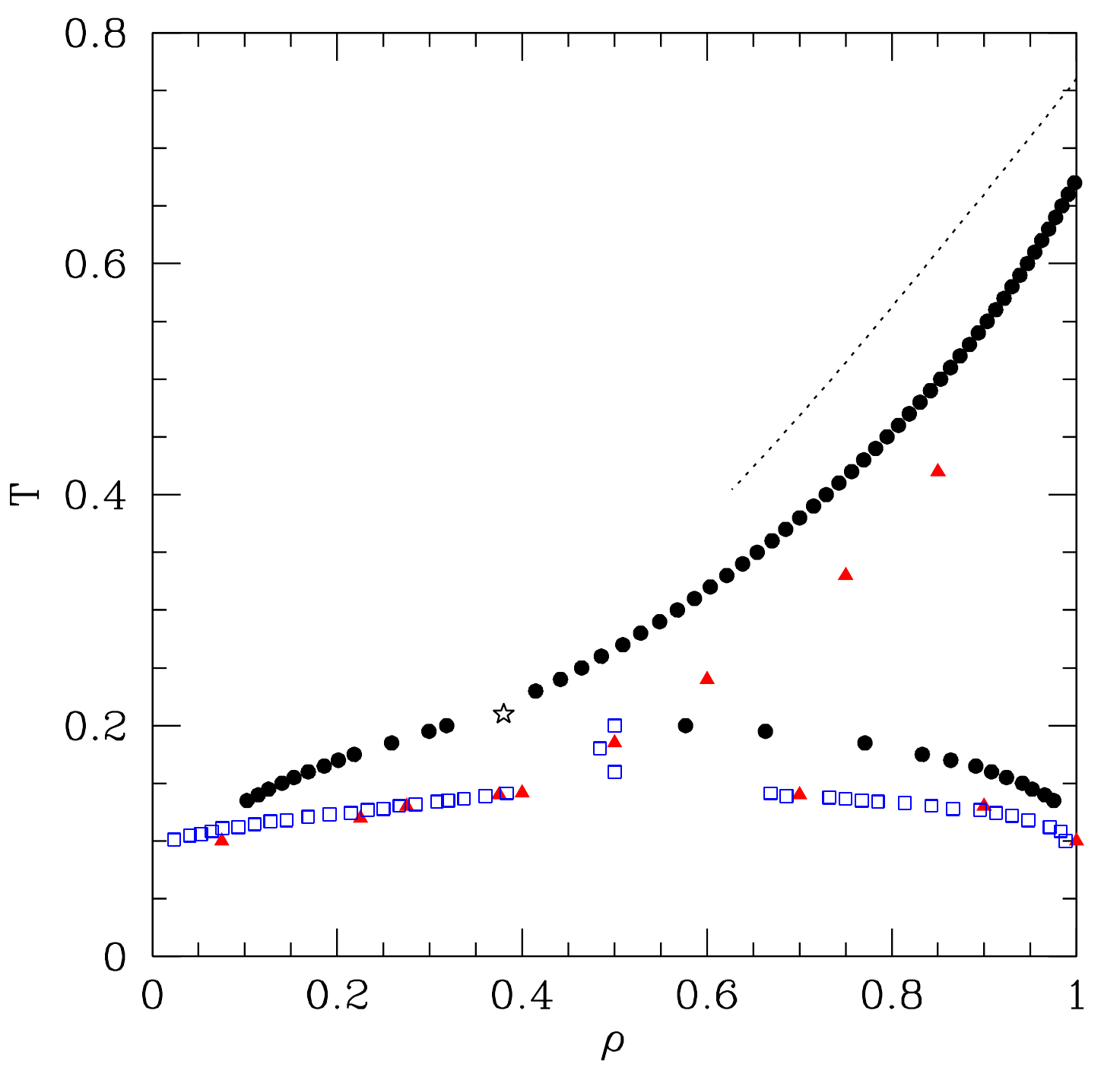}
\caption{Phase diagram of the LRPM in the $(\rho,T)$ plane. 
HRT results: the tricritical point is marked by an open star. 
Full dots give the $\lambda$-line above the tricritical temperature and the 
liquid-vapour binodal at lower temperatures.
The SCOZA data for the $\lambda$-line 
are shown by the dotted line. Simulation data from Refs. \cite{dickman,panagio} 
are represented by filled triangles and open squares respectively.
}
\label{figlrpm}
\end{figure}

In Ref. \cite{brognara2} the HRT formalism was generalized 
to the case of an uniaxial spin model where $S^z_\bR$ can assume three values: $-1,0,1$. This
system can be easily mapped into a symmetric binary mixture on a lattice with opposite signs for 
like and unlike interactions. If the coupling
is limited to nearest neighbours, the Blume-Capel model is obtained, while for long range
interactions decaying as $\sim 1/R$,
the Lattice Restricted Primitive Model (LRPM) is recovered. 
The phase diagrams of both models in the symmetry plane (zero magnetization) were
obtained by integrating the HRT flow equation \cite{brognara2}. 
The topologies of the phase boundaries in the density-temperature plane, 
shown in Fig. \ref{figlrpm} for the LRPM, turn out to be very similar:
at high temperatures a critical $\lambda$-line separates
a charge-ordered fluid phase at low density from a homogeneous plasma phase.
The critical line ends at a tricritical point where it merges into the liquid-vapour 
coexistence curve. The comparison with available simulation data for the LRPM shows 
that the structure of the phase diagram and the universality classes
of the transitions are correctly reproduced, 
although some quantitative discrepancy is present,
suggesting that a better representation of 
pair correlations should be devised. 

\subsection{Binary Mixtures}

The HRT formalism can be easily generalized to the case of binary mixtures \cite{mix}: 
the flow equations
in the smooth (\ref{smooth}) and sharp cut-off (\ref{sharp}) formulation remain valid for
mixtures provided the two body interaction $\phi(\bq)$ and the pair correlations 
${\cal F}_Q(\bq)$ and ${\cal C}_Q(\bq)$
are substituted by matrices whose indices run over the two species. 
The Ornstein-Zernike relation (\ref{calf}) acquires the form
\begin{equation}
\sum_j {\cal F}_Q^{ij}(\bq)\left [\frac{\delta^{jk}}{\rho_j}-\,{\cal C}_Q^{jk}(\bq)\right ] = 
\delta^{ik}
\label{calfmat}
\end{equation}
while the compressibility sum rule (\ref{komp}) becomes a set of three independent 
equations which can be written as:
\begin{equation}
\frac{\partial^2}{\partial\rho_i\partial\rho_j} \left (\frac{-\beta {\cal A}_Q)}{V}\right )
= {\cal C}_Q^{ij}(\bk=0) - \frac{\delta^{ij}}{\rho_i}
\label{kompmat}
\end{equation}
The derivation of the HRT evolution equations straightforwardly follows the 
same steps sketched in Section \ref{sec:exact}. By adopting the smooth cut-off 
procedure, the resulting expression is: 
\begin{equation}
\frac{1}{V}
\frac{\partial (-\beta {\cal A}_Q)}{\partial Q} = \frac{1}{2}\,\int \,\frac{{\rm d}\bq}{(2\pi)^d} \, 
{\rm Tr}\, \left \{ {\cal F}_Q(\bq) \left [ {\bf 1} + 
\left (\tilde \phi(\bq) - \tilde \phi_Q(\bq)\right )\, {\cal F}_Q(\bq)
\right ]^{-1} \, \frac{\partial \tilde\phi_Q(\bq)}{\partial Q} \right \}
\label{smoothmat}
\end{equation}
where matrix products are understood at right hand side and ``Tr" stands for the trace operation. 
In the sharp cut-off limit this form reduces to 
\begin{equation}
\frac{1}{V}
\frac{\partial (-\beta {\cal A}_Q)}{\partial Q} = - \frac{1}{2}\,\int_{q=Q} \,
\frac{{\rm d}\Omega_\bq}{(2\pi)^d} 
\, {\rm Tr}\,\log \left [ {\bf 1} + {\cal F}_Q(\bq)\tilde\phi(\bq)\right ]
\label{sharpmat}
\end{equation}
The trace of the logarithm at right hand side of Eq. (\ref{sharpmat})
must be interpreted in matrix sense and coincides with 
the logarithm of the determinant of the $2\times 2$ matrix 
$\left [{\bf 1} + {\cal F}_Q(\bq)\tilde\phi(\bq)\right ]$.
By use of a RPA-like closure of the form (\ref{rpa}), 
the flow equation then becomes a partial differential equation in three independent 
variables: the cut-off $Q$ and the densities of the two species $(\rho_1,\rho_2)$.

The analysis of criticality in binary mixtures 
is performed following the procedure developed in simple fluids \cite{adv,mix}: several 
universality classes are predicted by the HRT equations, but most of the possible fixed points
turn out to be unstable and the critical properties at a generic point along a transition 
line is of the Ising type. 
Actually, two distinct fixed points correspond to critical exponents 
related to the case of a one component scalar order parameter. One is weakly unstable and gives rise 
to strictly Ising-like critical properties, while the other is stable and 
correctly reproduces the so called phenomenological approach by Fisher which, starting from the
concept of field mixing, predicts the occurrence of some weak exponent renormalization \cite{phenom}: 
the long wavelength limit of a structure factor $S_{ij}(0)$
at the critical density and concentration diverges with an exponent 
$\gamma/(1-\alpha)$ (with $\alpha \simeq 0.11$), a $12\%$ larger than the corresponding 
value for one component fluids, while the isothermal compressibility displays 
a weak divergence governed by the exponent $\alpha/(1-\alpha)$.
HRT allows to identify the thermodynamic perturbation corresponding to the crossover between 
these two fixed points \cite{crox}, thereby providing a microscopic interpretation of the experimental 
difficulty in observing the true asymptotic critical behaviour. 
Special points in the phase diagram might
display genuinely different critical indices, like un-renormalized Ising or tricritical exponents. 

\begin{figure}
\includegraphics[height=7cm,width=7cm,angle=0]{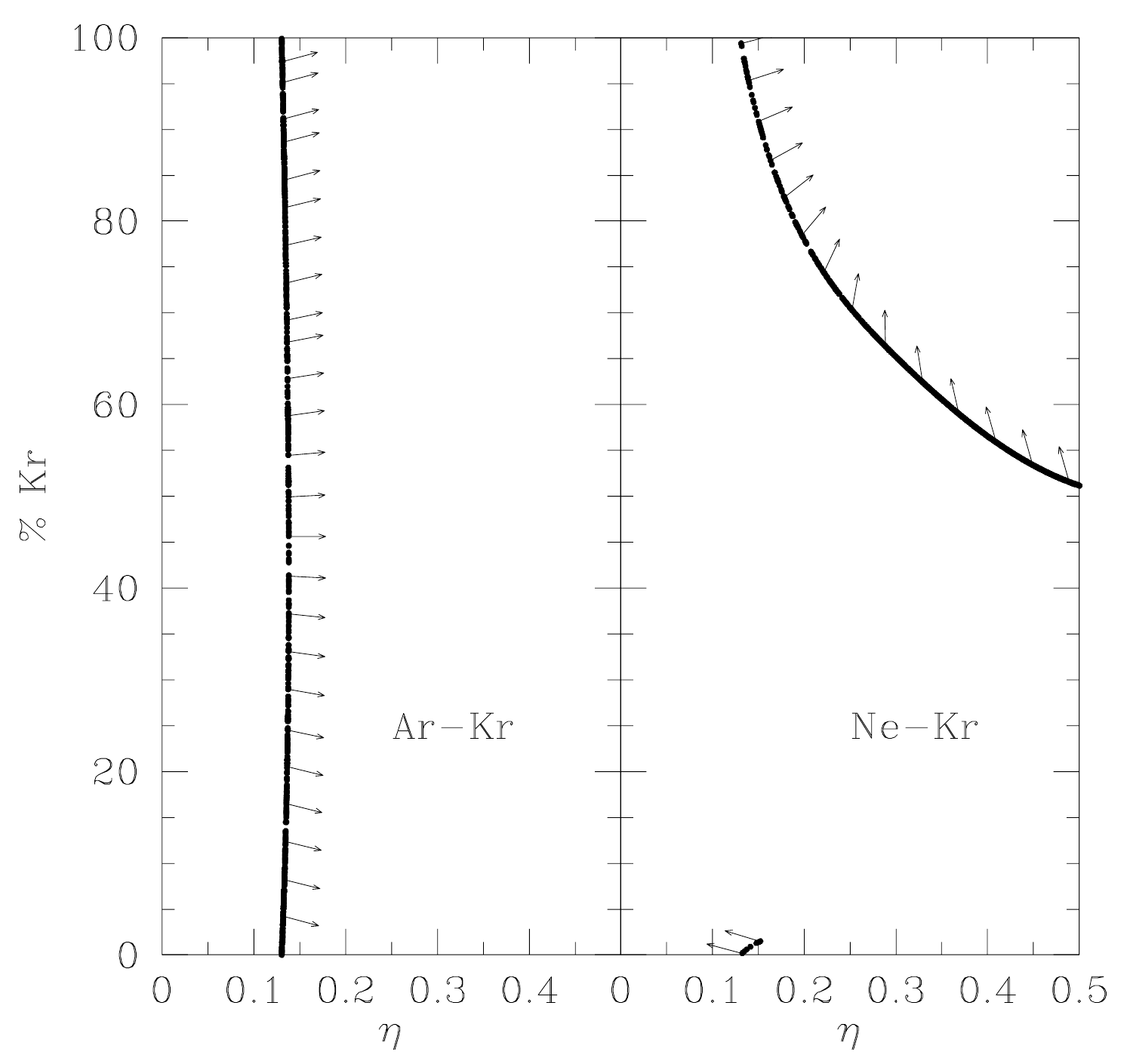}
\caption{Mean field critical lines projected 
in the plane total packing fraction/Krypton volume fraction,
for two parameter choices 
mimicking an Argon-Krypton (left panel) and a Nenon-Krypton mixture (right panel). The arrows
identify the mixing angle $\theta$ defined in the text: an arrow parallel to the density axis
indicates a liquid-vapour transition, an arrow along the volume fraction axis, a demixing transition.
}
\label{figmix}
\end{figure}

HRT also naturally suggests an unbiased definition of the order parameter at a critical
point of binary fluids \cite{mix}, which in general cannot be simply identified
either as density or concentration, because critical fluctuations are expected to 
affect both quantities. The critical point in binary fluids 
is signaled by the divergence of the leading eigenvalue of the (symmetric) matrix of the 
structure factors $S_{ij}(\bk)$ for $\bk\to 0$.
The structure factors are a direct measure of density fluctuations
and can be expressed as the equilibrium average of a product of density operators
$S_{ij}(\bk) = \frac{1}{N}\,\langle \hat\rho_i(\bk)\hat\rho_j(-\bk)\rangle$. 
To discriminate between demixing and liquid-vapour transitions in binary mixtures
it is convenient to introduce suitable linear combinations of $S_{ij}(\bk)$, 
the Bathia-Thornton structure factors \cite{bathia},
which properly characterize the density and concentration fluctuations: 
\begin{eqnarray}
S_{\rho\rho} &=& 
S_{11}(\bk)+S_{22}(\bk)+2\,S_{12}(\bk) \nonumber \\
S_{cc} &=&  
x_2^2\,S_{11}(\bk)+x_1^2\,S_{22}(\bk)-2\,x_1\,x_2\,S_{12}(\bk) \nonumber \\
S_{\rho c} &=& S_{c \rho} =
x_1\,S_{22}(\bk) - x_2\,S_{11}(\bk) + (x_1-x_2)\,S_{12}(\bk)
\end{eqnarray}
where $x_i=\rho_i/\rho$, $\rho=\rho_1+\rho_2$ are the number concentrations of the
two species and the total density, respectively. 
Let us denote by $\Lambda$ leading eigenvalue of the $2\times 2$ matrix 
built with the previously defined Bathia Thornton 
structure factors in the $\bk\to 0$ limit, 
and $(\cos\theta,\sin\theta)$ be the corresponding normalized eigenvector.
According to these definitions, the diverging eigenvalue provides a measure of the 
fluctuations of the particular linear combination of density and concentration
$\psi(\bk)=\cos\theta\,\rho(\bk) + \rho\,\sin\theta\,x_2(\bk)$ (for $\bk \to 0$): 
$\lim_{\bk\to 0}\,\frac{1}{N}\,\langle \psi(\bk)\psi(-\bk)\rangle=\Lambda$, 
suggesting the identification of
$\psi$ as the order parameter of the transition. It is convenient to pictorially
represent the order parameter as an arrow in the $(\rho,x_2)$ plane pointing in the
direction identified by $\theta$. Remarkably, the ``mixing angle" $\theta$ 
is related to well defined thermodynamic quantities \cite{adv}:
\begin{eqnarray}
1-x_2\,\cot\theta &=&\rho\,{\rm v}_1  \label{termo1}\\
\left [\frac{\Delta x_2}{\Delta (1/\rho)} \right ]_{coex}&=&\rho\tan\theta
\quad .
\label{termo2}
\end{eqnarray}
The first equation is written in terms of the partial 
molar volumes ${\rm v}_i=(\partial V/\partial N_i)_{P,T}$
which attain a finite limit at the critical point. The second, equivalent, relation 
expresses $\theta$ in terms of the differences of concentration $\Delta x_2$
and molar volumes $\,\,\Delta (1/\rho)$ at coexistence on an isotherm--isobar
near the critical point. Clearly, the direction of the order parameter is defined at each point of
a critical line and varies continuously along that line. While in some mixture, the character 
of the transition remains almost unaltered by varying the thermodynamic state, in other cases it may
instead experience considerable changes, passing from demixing to liquid-vapour by reducing the density. 
A plot of the mean field critical lines for two Lennard-Jones binary fluids 
mimicking two rare gas mixtures, shown in Fig. \ref{figmix}, illustrates these two possibilities. 
For convenience, the total packing fraction $\eta=\frac{\pi}{6}(\rho_1 \sigma_1^3 + \rho_2 \sigma_2^3)$
and the volume fraction $\eta_2/\eta$ are 
used in place of total density $\rho$ and number concentration $x_2$. 

The numerical integration of the HRT equation poses some 
challenge due to the presence of three independent 
variables: cut-off $Q$, density $\rho$ and concentration $x_2$. Special attention must be paid to 
the stability requirements and to the possible occurrence of singularities induced by the 
presence of large coexistence regions in the phase diagram. 
A stable numerical code has been developed for two classes of models of binary fluids: 
the rare gas mixtures \cite{mixlj} and the symmetric Yukawa mixtures \cite{mixyuk}, 
which include the Ising fluids, defined as hard spheres 
with an internal Ising-like internal degree of freedom whose relative 
orientation specifies the sign of the
pair interaction \cite{isfl}. In both cases, the HRT equation in the sharp cut-off formulation
has been numerically integrated. 
Here we just review few results for the model of rare gas mixtures, where 
extensive experimental data are available \cite{shouten}.  

\begin{figure}
\includegraphics[height=6cm,width=6cm,angle=0]{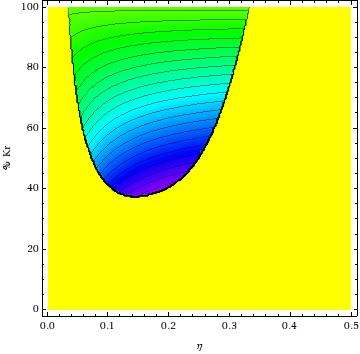}
\caption{Coexistence regions in the packing fraction/Krypton molar fraction in the
Argon-Krypton mixture according to HRT at $T=-95.77^\circ C$. 
Tie lines in the two-phase regions are shown. 
In the coexistence regions darker colors correspond to higher pressures.
}
\label{figarkr1-rx}
\end{figure}

\begin{figure}
\includegraphics[height=6cm,width=6cm,angle=0]{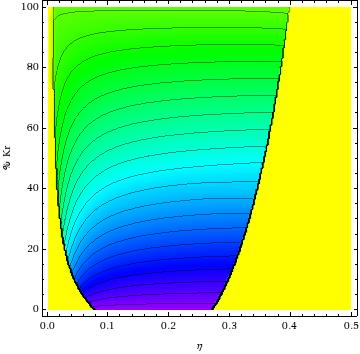}
\caption{Coexistence regions in the packing fraction/Krypton molar fraction in the
Argon-Krypton mixture according to HRT at $T=-130^\circ C$.
Tie lines in the two-phase regions are shown.
In the coexistence regions darker colors correspond to higher pressures.
}
\label{figarkr2-rx}
\end{figure}

\begin{figure}
\includegraphics[height=7cm,width=7cm,angle=0]{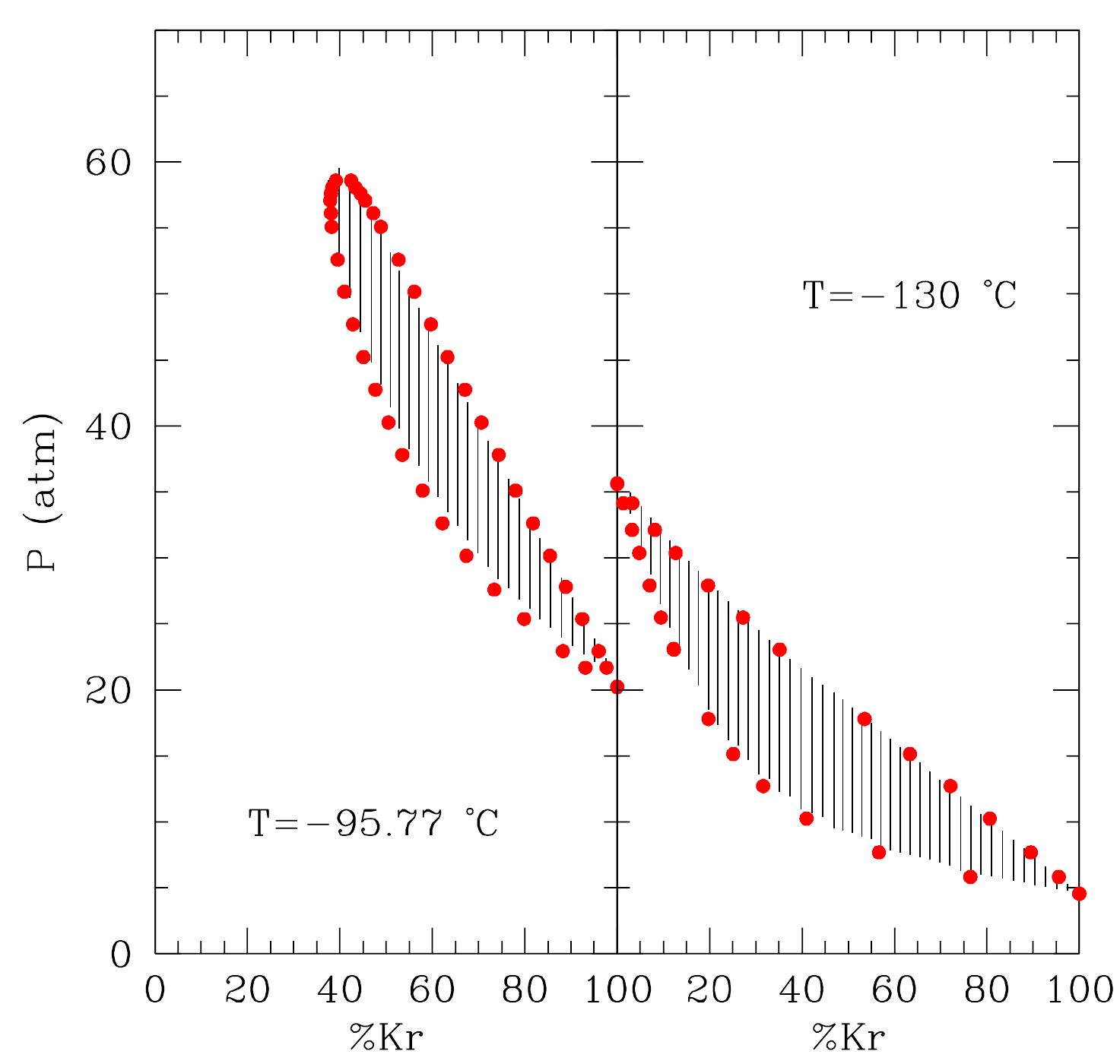}
\caption{Coexistence regions in the Krypton molar fraction/pressure plane for the
Argon-Krypton mixture according to HRT. Left panel: $T=-95.77 ^\circ C$, right panel
$T=-130 ^\circ C$. Experimental phase boundaries \cite{arkr} are shown as red dots.
}
\label{figarkr-xp}
\end{figure}

\begin{figure}
\includegraphics[height=7cm,width=7cm,angle=0]{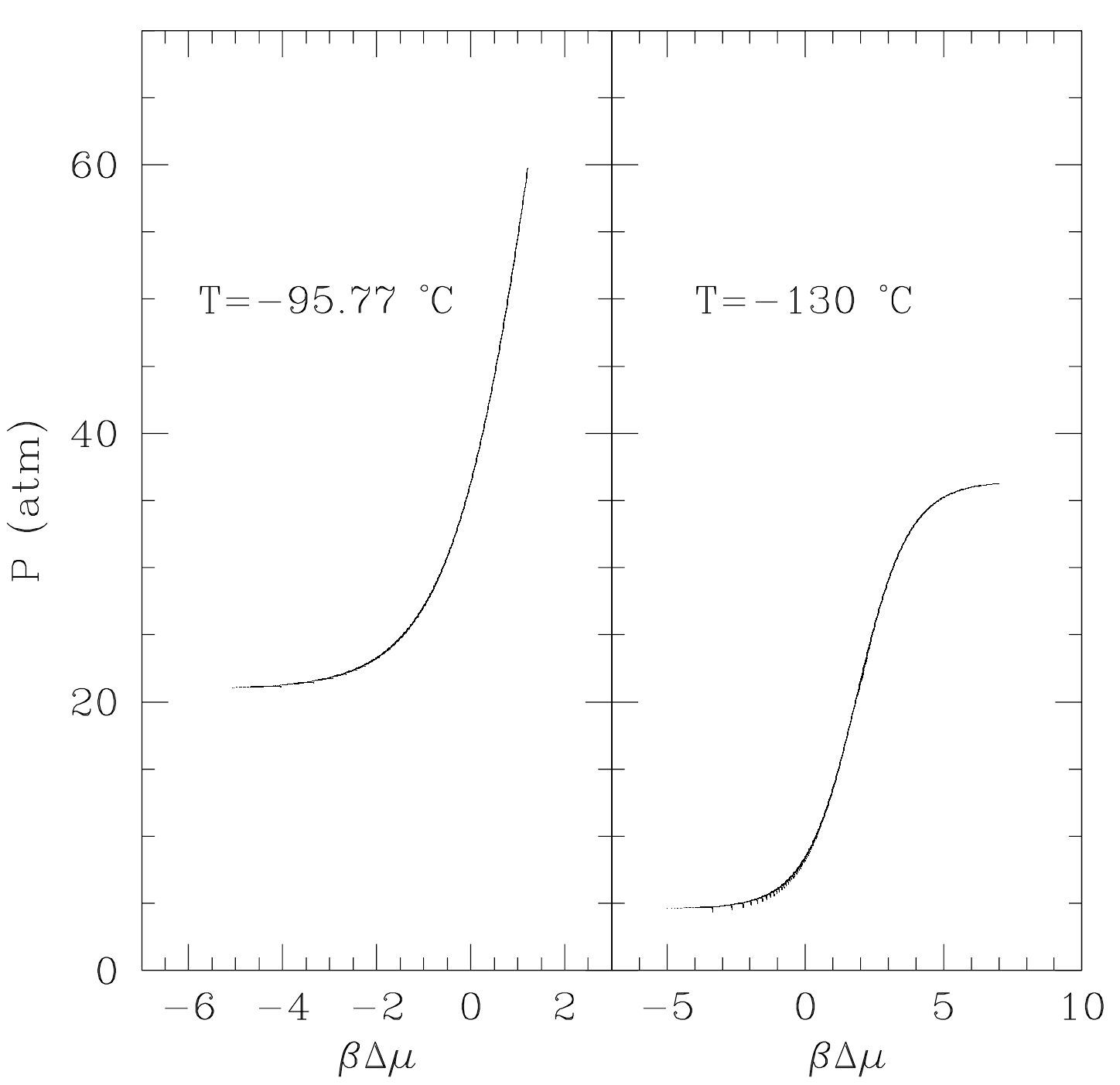}
\caption{Coexistence regions for the Argon-Krypton mixture according to HRT plotted in
the plane dimensionless chemical potential difference/pressure. 
Left panel: $T=-95.77 ^\circ C$, right panel $T=-130 ^\circ C$. Note the collapse of the
full coexistence region into a single line. 
}
\label{figarkr-mup}
\end{figure}

The Argon-Krypton system is modeled as a mixture of particles interacting via 
Lennard-Jones (LJ) potentials \cite{mixlj}. 
Each LJ interaction has been preliminarily mapped into a reference 
hard sphere potential plus a LJ attractive tail. The effective
hard sphere diameter does not coincide with the LJ parameter $\sigma$ and is determined 
according to the standard Barker-Henderson prescription \cite{hansen} which gives rise to a weak
temperature dependence.  
The bare LJ parameters adopted in the numerical calculation have been fixed to
reproduce the location of the critical points of the two pure components: 
$\sigma_{\rm Ar}=3.52$ \AA, $\sigma_{\rm Kr}=3.77$ \AA, 
$\epsilon_{\rm Ar-Ar}=113.8$ K and $\epsilon_{\rm Kr-Kr}=157.9$ K. The strength of the unlike attractive 
interaction has been set to $\epsilon_{\rm Ar-Kr}=133.5$ K, and additivity of the 
effective hard sphere diameters has been assumed. Plots of the coexistence regions in the 
packing fraction/molar concentration plane are given 
for two temperatures in Figs. \ref{figarkr1-rx}-\ref{figarkr2-rx}, where 
the tie-lines, connecting thermodynamic states in equilibrium, are also shown. The Argon-Krypton
mixture belongs to the topology of phase diagram illustrated in the left panel of Fig \ref{figmix},
where the critical points of the two species are connected by a line of critical points. 
A different view of the coexistence regions at the same temperatures
can be seen in the molar concentration/pressure plane 
in Fig. \ref{figarkr-xp}: at the higher temperature pure Argon is in the fluid 
phase, while at the lower temperature the coexistence regions of the two components 
merge and define a unique, connected domain. 
A good agreement between theory and experiments \cite{arkr} can be appreciated for this system. 
Finally, it is instructive to check that, within HRT, the full coexistence region collapses to 
a single line when intensive thermodynamic quantities are selected, as required by 
thermodynamics: in Fig. \ref{figarkr-mup} 
the same coexistence regions are plotted in a diagram where the difference between 
the chemical potentials of the
two species and the pressure are considered as independent variables.  

\begin{figure}
\includegraphics[height=6cm,width=6cm,angle=0]{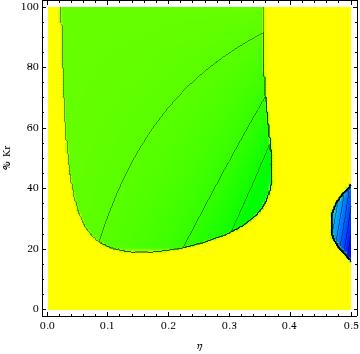}
\caption{Coexistence regions in the packing fraction/Krypton molar fraction in the
Neon-Krypton mixture according to HRT at $T=-110^\circ C$.
Tie lines in the two-phase regions are shown.
In the coexistence regions darker colors correspond to higher pressures.
}
\label{fignekr1-rx}
\end{figure}

\begin{figure}
\includegraphics[height=6cm,width=6cm,angle=0]{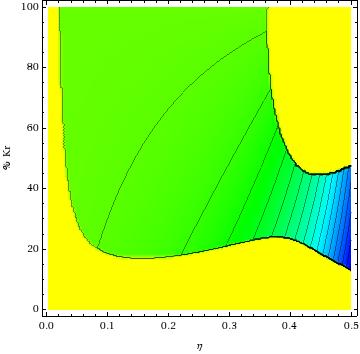}
\caption{Coexistence regions in the packing fraction/Krypton molar fraction in the
Neon-Krypton mixture according to HRT at $T=-107^\circ C$.
Tie lines in the two-phase regions are shown.
In the coexistence regions darker colors correspond to higher pressures.
}
\label{fignekr2-rx}
\end{figure}

\begin{figure}
\includegraphics[height=7cm,width=7cm,angle=0]{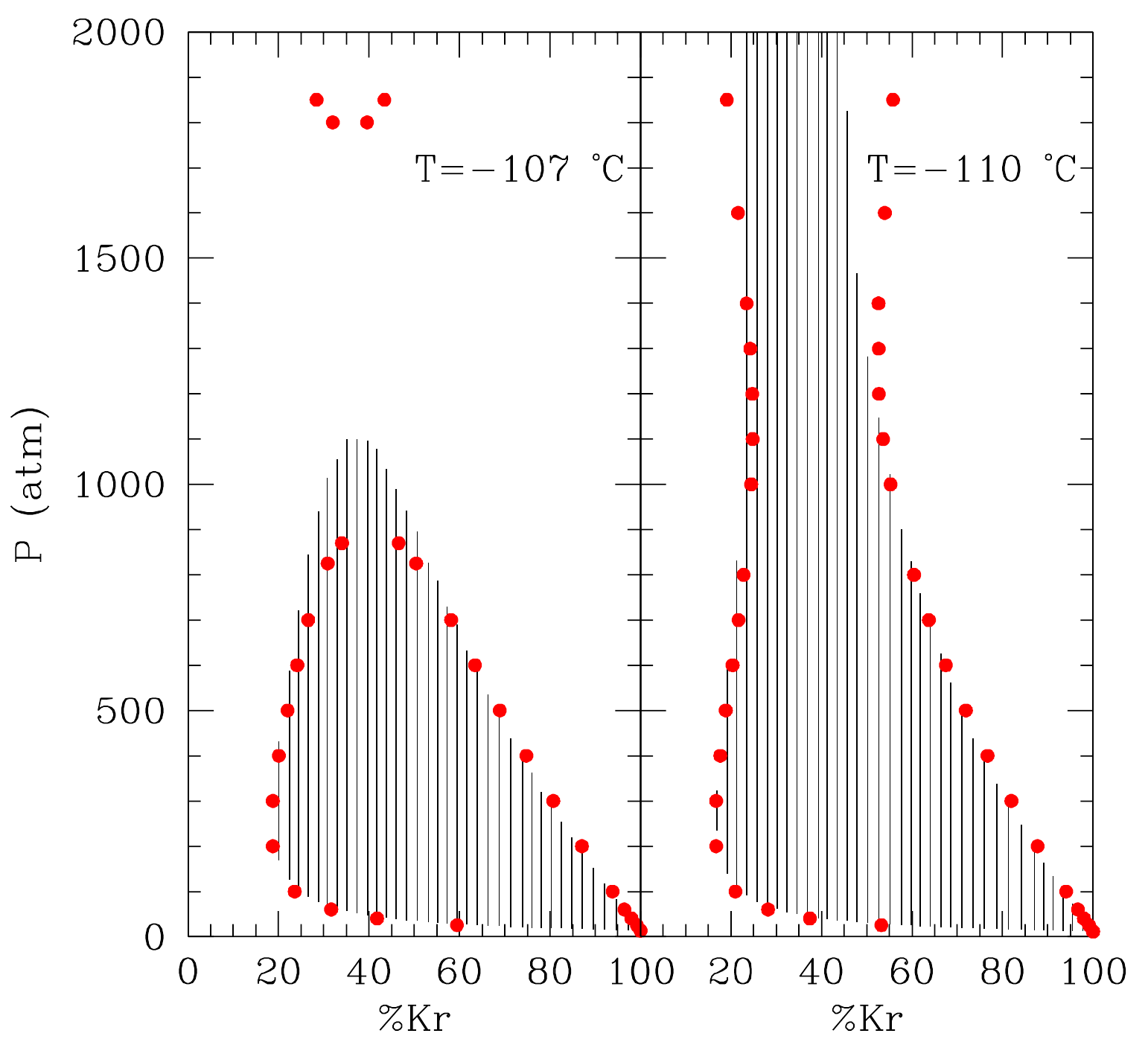}
\caption{Coexistence regions in the Krypton molar fraction/pressure plane for the 
Neon-Krypton mixture according to HRT. Left panel: $T=-110 ^\circ C$, right panel 
$T=-107 ^\circ C$. Experimental phase boundaries \cite{nekr} are shown as red dots.
}
\label{fignekr-xp}
\end{figure}

\begin{figure}
\includegraphics[height=7cm,width=7cm,angle=0]{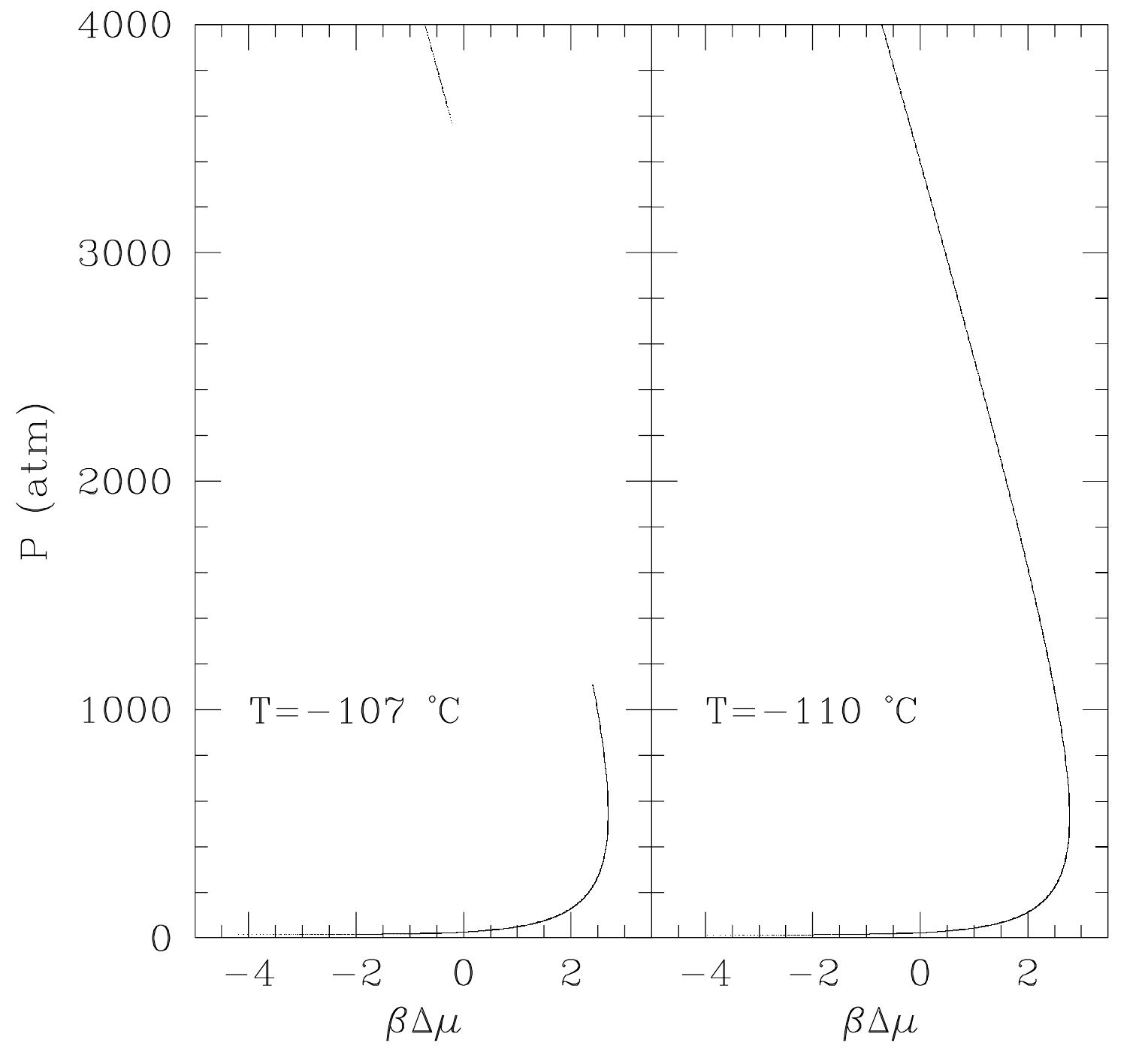}
\caption{Coexistence regions for the Neon-Krypton mixture according to HRT plotted in
the plane dimensionless chemical potential difference/pressure.
Left panel: $T=-107 ^\circ C$, right panel $T=-110 ^\circ C$. 
}
\label{fignekr-mup}
\end{figure}

The phase diagram of the Neon-Krypton mixture instead displays a different 
topology, belonging to the class depicted in the right panel of Fig. \ref{figmix}. 
The same Lennard-Jones model used for the Argon-Krypton mixture has been adopted with 
parameters appropriate for the Neon-Neon ($\sigma_{\rm Ne}=2.85$ \AA, 
$\epsilon_{\rm Ne-Ne}=33.5$ K) and Neon-Krypton interactions ($\epsilon_{\rm Ne-Kr}=54.4$ K).
The coexistence regions of the two pure components 
remain separated in the whole phase diagram: while pure Neon is in the gas phase in a large 
temperature range, the Neon-Krypton mixture displays phase coexistence between states of 
considerably different composition. The order parameter smoothly changes character
signaling a liquid-vapour transition at low Neon concentration and a demixing transition 
at higher pressures as already captured by the mean field calculations of Fig \ref{figmix}. 
The coexistence region in the three planes packing fraction/molar concentration, 
molar concentration/pressure and chemical potential/pressure are shown in Figs. \ref{fignekr1-rx},
\ref{fignekr2-rx}, \ref{fignekr-xp} and \ref{fignekr-mup} respectively. 
The comparison with the experimental data \cite{nekr} shows good agreement at not too high pressure, 
while some discrepancy is present at high pressures, where the repulsive part 
of the interaction plays an important role. As discussed in Ref. \cite{mixlj}, it is likely that
a better representation of the soft core part of the interatomic potential 
must be employed for a fully quantitative description of demixing transitions at high pressures.

\section{HRT for Quantum systems}

\begin{figure}
\includegraphics[height=7cm,width=7cm,angle=0]{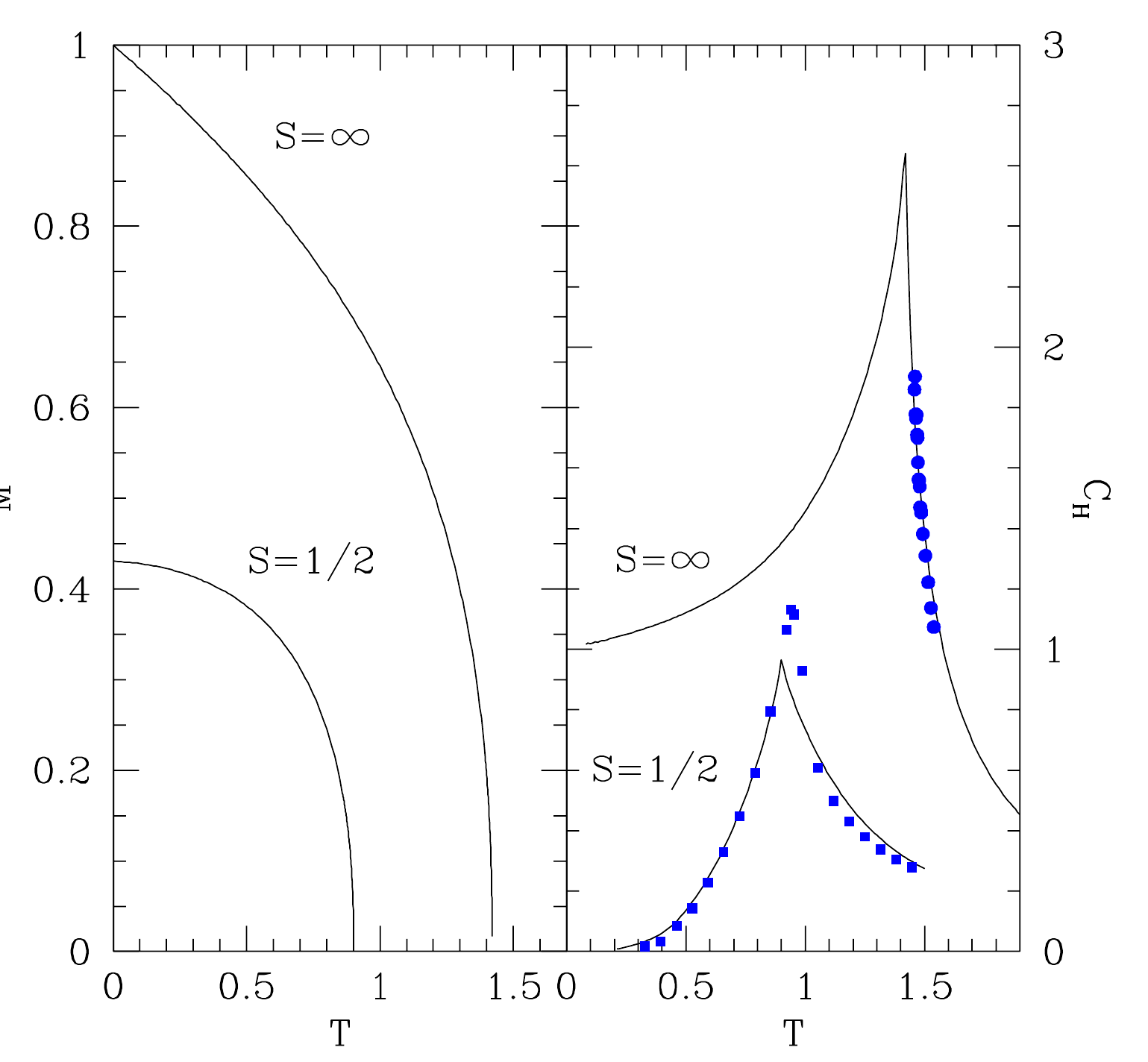}
\caption{QHRT results for the antiferromagnetic Heisenberg model in $d=3$ 
for spin $S=\infty$ and $S=1/2$.
Left panel: QHRT spontaneous magnetization for $S=1/2$ and $S=\infty$ (classical limit). 
Right panel: constant field specific heat at $h=0$ as a function of temperature compared with 
classical (full dots) \cite{holm} and quantum (full squares) \cite{sandvik} Monte Carlo 
simulations, respectively. 
In the $S=\infty$ limit the magnetization is normalized to $S$ and the specific heat to $S^2$.
}
\label{figheis}
\end{figure}

HRT was originally developed in the framework of classical statistical physics
but its formalism can be generalized to quantum hamiltonians. Lattice spin models 
have been considered in Ref. \cite{gianinetti}, where the reference system was
identified as free spins in an external magnetic field which couples with the order parameter. 
Quantum antiferromagnets on an hypercubic lattice in $d$ dimensions are 
described by the isotropic, nearest neighbour Heisenberg hamiltonian
\begin{equation}
H = J\, \sum_{\langle \bR,\bR^\prime\rangle} \,\bS_\bR \cdot \bS_{\bR^\prime} - h_s\sum_R 
e^{i\bg\cdot\bR} S^z_\bR
\label{heis}
\end{equation}
and require the introduction of a staggered external field $h_s$.
Here, $\bS_\bR$ are spin $S$ operators defined on the sites $\bR$ of the lattice and
$\bg=(\pi,\pi\dots)$ is the antiferromagnetic vector in $d$ dimensions.  
The reference system is defined by the hamiltonian (\ref{heis}) for $J=0$ and gives rise to 
a solvable, single site problem. Following the standard sharp cut-off formulation of HRT, 
the spin-spin interaction is gradually turned on, by 
including its Fourier components starting from the boundary of the
antiferromagnetic Brillouin Zone. The exact evolution equation for the free energy per site 
in sharp cut-off formulation was 
obtained in Ref. \cite{gianinetti} and involves the two body dynamical spin-spin correlations 
in imaginary time: 
\begin{eqnarray}
\frac{1}{N}\frac{\partial{\cal A}_Q}{\partial Q} &=& 
\frac{kT}{2}\, \sum_{\omega_n}\, \int_{BZ} \frac{{\rm d}\bk}{(2\pi)^d}\,
\delta\left ( \sqrt{\tilde J(0)^2 -\tilde J(\bk)^2} - Q \right ) \, \Big \{
\log \left [ 1-{\cal F}_Q^{zz}(\bk,\omega_n) \tilde J(\bk)\right ] + \\
&& \log \left [ \left (1-{\cal F}_Q^{xx}(\bk,\omega_n) \tilde J(\bk)\right ) \,
\left ( 1-{\cal F}_Q^{xx}(\bk+\bg,\omega_n) \tilde J(\bk+\bg)\right ) + 
{\cal F}_Q^{xy}(\bk,\omega_n)\,{\cal F}_Q^{xy}(\bk+\bg,\omega_n)\,
\tilde J(\bk)\tilde J(\bk+\bg) \right ] \Big \} \nonumber 
\label{gian}
\end{eqnarray} 
where $\tilde J(\bk)$ is the Fourier transform of the spin-spin nearest negighbour interaction
and $Q$ ranges from $\tilde J(0)$, where fluctuations are absent and the 
mean field approximation is
recovered, to $Q=0$, where all Fourier components are introduced. 
The structure of this equation is rather similar to its classical counterpart (\ref{lsharp})
with two main differences: the presence of off diagonal two body correlations 
${\cal F}_Q^{xy}(\bk,\omega_n)$, due to the non commutativity of the spin components,  
and the summation over the Matsubara frequencies $\omega_n=2\pi\,kT\,n$. 
The imaginary time variable in fact naturally emerges when quantum perturbation theory
is employed to evaluate the effects of an additional infinitesimal momentum shell in the 
spin-spin interaction. Notice that in this formulation, the cut-off on fluctuations is 
set only on the spatial Fourier components, while the Matsubara frequency variable 
runs on the whole real axis for all $Q$'s. The Quantum HRT (QHRT) equation 
appropriate for the antiferromagnetic Heisenberg model (\ref{heis}) was studied 
by employing a simple closure similar in spirit to the classical RPA-like approximation
(\ref{lrpa}): the structure of the dynamical correlations was evaluated at mean field level
and only ``mass renormalization" was allowed by imposing thermodynamic consistency between the 
zero frequency limit of the dynamical correlations at wave-vector $\bk=\bg$
and the derivatives of the free energy 
${\cal A}_Q$ with respect to the staggered magnetization. The explicit form of the adopted closure 
reads: 
\begin{eqnarray}
{\cal F}^{xx}_Q(\bk,\omega) &=& 
\frac{\mu_\perp -\tilde J(\bk)}{m_s^{-2}\omega^2 +\mu_\perp^2 - \tilde J(\bk)^2} \\
{\cal F}^{xy}_Q(\bk,\omega) &=& 
\frac{m_s^{-1}\omega}{m_s^{-2}\omega^2 +\mu_\perp^2 - \tilde J(\bk)^2} \\
{\cal F}^{zz}_Q(\bk,\omega) &=& 
\frac{\delta_{\omega,0}}{\mu_\parallel + \tilde J(\bk)}
\end{eqnarray}
where $m_s$ is the staggered magnetization, and the two cut-off dependent 
parameters $(\mu_\perp,\mu_\parallel)$ 
are related to the transverse and longitudinal staggered susceptibilities:
\begin{eqnarray}
\frac{\partial}{\partial m_s} \left (\frac{{\cal A}_Q}{N}\right )
&=& m_s\,\left [\mu_\perp + \tilde J(\bg) \right ]\\
\frac{\partial^2}{\partial m_s^2} \left (\frac{{\cal A}_Q}{N}\right )
&=& \mu_\parallel + \tilde J(\bg) 
\end{eqnarray}
In Fig. \ref{figheis} the constant field specific heat and the 
spontaneous magnetization of a Heisenberg antiferromagnet on the cubic lattice
below the critical temperature are shown for two choices of the spin $S$: the extreme quantum 
value $S=1/2$ and the classical limit $S\to\infty$. 
Monte Carlo simulations for $C_H$ \cite{holm,sandvik} are also reported. 
A more detailed comparison between theory and simulations 
can be found in \cite{gianinetti}, showing that 
the performances of QHRT have generally a lower quality with respect to the 
previously shown applications to classical models, suggesting that a better representation
of static and dynamic spin correlations has to be developed: the different role of longitudinal and 
transverse correlations is probably not fully grasped by the adopted single mode approximation,
which did not prove accurate enough to provide a consistent picture of the 
crossover between the renormalized classical and the quantum critical regime of the 
Non Linear Sigma Model \cite{chn}. 

The effects of easy-axis anisotropy have been also investigated by use of QHRT \cite{gianinetti2}, as well
as the role of a weak next nearest neighbour interaction \cite{spanu}. However, a QHRT analysis 
of quantum (i.e. zero temperature) phase transitions, or of the occurrence of off-diagonal long range 
order has not been attempted yet. As shown in the related non perturbative RG approach for the 
Bose-Hubbard model \cite{dupuis2}, the investigation of Bose condensation by RG methods requires 
a different cut-off procedure and it may be more convenient to set a cut-off on the kinetic 
energy term rather than on the two body interaction. According to this scheme, 
the reference system becomes a purely classical model, while quantum fluctuations are gradually
introduced, leading to the appropriate free energy flow equation. 

The smooth cut-off formulation of HRT has not been applied to quantum systems yet, although the
derivation of the flow equations does not present any additional difficulty. A closer connection
between QHRT and Non Perturbative Renormalization Group methods may provide some clue on
the structure of more refined closures to the flow equation.

\section{Conclusions}

A proper description of first and second order phase transitions within liquid state theory
requires a careful account of long wavelength fluctuations. The Hierarchical Reference Theory 
of fluids was developed precisely to achieve this goal. While the general framework of the theory
was set since the very first formulation of HRT, the implementations to specific physical systems 
required the development of approximation schemes which may vary according to the individual 
features of the model. In this review we tried to summarize the efforts in this direction 
carried out in the last ten years. 
Among the features shared by all the non-perturbative closures of the HRT
equation we may mention the correct scaling form of the thermodynamics and correlations in the 
critical region, the prediction of non-classical critical exponents and the exact implementation
of the Maxwell construction at first order transitions, showing that 
the inclusion of long wave-length fluctuations
gives rise to coexistence regions at constant pressure and chemical potential. The latter feature 
is particularly valuable in the case of mixtures, where the {\it ad hoc} implementation
of the equal tangent construction, required by most liquid state theories, may prove rather cumbersome. 
The smooth cut-off formulation of HRT, proposed several years ago, has been quantitatively 
investigated only recently. Among its peculiar properties, it is worth mentioning the 
absence of the spurious divergence of the isothermal compressibility at the phase boundary:
a problem which affects the sharp cut-off formulation. This allows the location of the spinodal loci
within a Renormalization Group approach, as explictly shown for the Yukawa fluid.  
Extensions of the smooth cut-off HRT equations to 
the case of mixtures or quantum systems are straightforward and represent a fertile 
research field for future studies. 

The approximations to the HRT hierarchy we have been considering so far 
were limited to closures to the first equation, which provides the flow equation for the 
free energy of the model. More sophisticated closures, at the level of the second equation, 
which contains information on the two particle correlations, have been formulated but not 
quantitatively investigated yet. 

Soft matter physics provides a stimulating environment 
for any microscopic theory able to capture the mechanisms leading to phase transitions. 
And HRT was applied to several systems of physical interest in this context. However, we feel 
that improved analytical representations of pair correlations in different 
frameworks have still to be developed to achieve a degree of accuracy comparable to 
simulation methods. For instance, a common feature of colloidal suspensions is the 
presence of short range attractive interactions, often represented by the ``sticky limit". 
A closure of the form (\ref{rpa}), adopted in most of the implementations of HRT, 
implies a linear relation between the direct correlation function and the attractive 
part of the potential, which is known to provide poor results for short range interactions \cite{hansen}. 
Extensions of HRT to lattice systems, mixtures and quantum models have 
been briefly covered in this review, but other interesting directions were explored in the literature, 
as the application of HRT to polymer physics or to disordered media \cite{tarjus}. 

A special mention must be devoted to the relationship between HRT and non perturbative 
(or functional) Renormalization Group methods. The two approaches are in fact intimately related, 
as already noticed by Caillol \cite{caillol},
and have been independently developed in different physical frameworks.
We hope that the present review,
which specifically devoted Section \ref{sec:nprg} to this problem, 
will contribute to clarify this issue providing a dictionary between the two, substantially equivalent,
formulations. 

It is a pleasure to express our gratitude to 
all the former students and collaborators who 
helped us in these investigations,
most notably  D. Pini, M. Tau and F. Lo Verso. 
We acknowledge financial support by the 
CIMAINA foundation (LR) and by MiUR through a PRIN/08 grant (AP). 

\section{Appendix}

The flow equation for the free energy (\ref{sharp}), 
together with the definition (\ref{calf}) and the parametrization
of the direct correlation function (\ref{rpa}), supplemented by the compressibility sum rule (\ref{komp})
form a closed partial differential equation (PDE) for the free energy as a function 
of the density $\rho$ and the cut-off $Q$. As a practical example here we report the explicit equations 
for a model defined by a spherically symmetric interaction written as the sum of a 
reference potential $v_R(r)$ (and corresponding free energy $A_{R}(\rho)$) 
and an attractive tail $w(r)$. As usual we set $\phi(r)=-\beta w(r)$; Fourier transforms are denoted 
by tildes. 

The initial condition of the PDE is set at some large value of $Q_\infty$, such that the
pair interaction $\tilde w(\bq)$ can be neglected for $q>Q_\infty$. 
The free energy at $Q_\infty$ is then given by the mean field approximation:
\begin{equation}
{\cal A}_{\infty}(\rho) = A_{R}(\rho) + V\,\frac{\rho^2}{2} \tilde w(0)
\end{equation}
Here and in the following, the ideal gas contribution is included in the definition of $A_R$.
The integration of the PDE is then performed backwards, down to $Q=0$, where 
all fluctuations are included and ${\cal A}_Q$ coincides with the free energy of the model.
A mesh is defined on the density axis $(0,\rho_{max})$. Here $\rho_{max}$ is some relatively 
high value (typically $\rho_{max}\sim 1$ in units of the hard-sphere diameter) and the free energy
${\cal A}_Q(\rho)$ is propagated from $Q$ to $Q-dQ$ by discretizing the PDE on the mesh. 
It is convenient to write the PDE (\ref{sharp}) in quasi-linear form by introducing the 
auxiliary quantity
\begin{equation}
u_Q(\rho) = \log \left [ 1 + {\cal F}_Q(Q)\,\tilde \phi(Q) \right ]
\label{defu}
\end{equation}
The evolution equation for this new variable can be straightforwardly obtained from Eq. (\ref{sharp}):
\begin{equation}
\frac{\rho\,\tilde \phi(Q)\,e^{u_Q}}{\left (e^{u_Q}-1\right )^2}\,
\frac{\partial u_Q}{\partial Q} = 
\rho\,\frac{\partial \tilde c_R(Q)}{\partial Q} +
\frac{1}{\tilde\phi(Q)}\,\frac{\partial \tilde \phi(Q)}{\partial Q} 
\left [1 - \rho\,\tilde c_R(Q) \right ] -\rho\,Q^{d-1}\,
\frac{\tilde\phi(Q)}{\tilde\phi(0)}\,\frac{d\,K_d}{2} \,\frac{\partial^2 u_Q}{\partial\rho^2}
\label{diff}
\end{equation}
where as usual $K_d=(6\pi^2)^{-1}$ in three dimensions ($d=3$) and $\tilde c_R(Q)$ is the direct
correlation function of the reference system. Usually the hard sphere gas is adopted as a reference 
system and $\tilde c_R(Q)$ is evaluated in Percus-Yevick \cite{hansen}
or Verlet-Weis approximation \cite{verlet}.
This PDE has the structure of a non-linear diffusion equation, where $-Q$ plays the role of
time-like variable and $\rho$ of space-like variable. The stability requirement for 
this class of PDE's is given by the positiveness of the effective diffusion coefficient. 
This condition is verified by Eq. (\ref{diff}) provided $\tilde\phi(0)$ is positive. 

Boundary conditions at $\rho=0$ and $\rho=\rho_{max}$ are required: it 
follows from the definition that at low density $u_Q(\rho) \to \rho\,\tilde\phi(Q)$, while
$u_Q(\rho_{max})$ is usually approximated by its mean field expression, i.e. by Eq. (\ref{defu})
with ${\cal F}_Q(Q)=\rho/[1-\rho\,\tilde c_R(Q)-\rho\,\tilde\phi(Q)]$. Different boundary 
conditions at high density give very similar solutions to the PDE, at least when the range
of the potential is not too short.  

Details on the discretization procedure can be found in Refs. \cite{meroni,tau}.
We finally remark that at low values of $Q$ (e.g. $Q<Q_0\sim 0.1$) it is convenient to adopt a
logarithmic scale in the cut-off, by setting $Q=Q_0\,e^{-t}$ with $t\in (0,\infty)$ to slow down
the inclusion of long wave-length fluctuations.


\begin{thebibliography}{10}

\bibitem{hansen}
J.P.~Hansen and I.R.~McDonald,
{\em Theory of Simple Liquids}, 3$^{\rm rd}$ ed., Academic Press,
London (2006).

\bibitem{barrat} 
J.L. Barrat and J.P. Hansen, 
{\em Basic concepts for simple and complex liquids}, Cambridge University Press (2003).

\bibitem{rg}
K.G.~Wilson and J.B.~Kogut, Phys. Rep. C {\bf 12}, 75 (1974).

\bibitem{hrt0}
A.~Parola and L.~Reatto,
Phys. Rev. Lett. {\bf 53}, 2417 (1984); Phys.Rev. A {\bf 31}, 3309 (1985).

\bibitem{adv}
A.~Parola and L.~Reatto, Adv. Phys. {\bf 44}, 211 (1995).

\bibitem{wetterich} J. Berges, N. Tetradis and C. Wetterich, Phys. Rep. {\bf 363} 223 (2002).

\bibitem{caillol} J.M. Caillol, Mol. Phys. {\bf 104}, 1931, (2006); {\it ibid.} {\bf 109},
2813 (2011).  

\bibitem{note1} Note that the diagrammatic resummation quoted in Eq. (4.17) of Ref. \cite{adv} is
reported incorrectly. The expression ``$\Delta$ bond" should be substituted with ``$F_2^R$ bond"
throughout and the sentence ``exactly one path" in item (ii) should read ``at most one path". 

\bibitem{delamotte} B. Delamotte, 
``An Introduction to the Nonperturbative Renormalization Group", 
http://xxx.lanl.gov/abs/cond-mat/0702365v1.

\bibitem{ap}
A.~Parola, J. Phys. C {\bf 26}, 5071 (1986).

\bibitem{morris} T. Morris, Int. J. Mod. Phys. A {\bf 9}, 2411 (1994). 

\bibitem{nicoll} J.F. Nicoll, T.S. Chang, H.E. Stanley, Phys. Rev. Lett.
{\bf 33}, 540 (1974) and Phys. Rev A. {\bf 13}, 1251 (1976);
J.F. Nicoll, T.S. Chang, Phys. Lett. {\bf 62A}, 287 (1977).

\bibitem{hubbard} J. Hubbard, P. Schofield, Phys. Lett. {\bf 40A}, 245 (1972).

\bibitem{dupuis} T. Machado and N. Dupuis, \pre {\bf 82}, 041128 (2010). 

\bibitem{tarazona} P. Tarazona and R. Evans, Mol. Phys. {\bf 52}, 847 (1984). 

\bibitem{note3} Here we assume that the regular part of the potential has negative definite 
Fourier transform $\tilde w(\bq) < 0$. 

\bibitem{smooth} A. Parola, D. Pini, L. Reatto, Mol. Phys.  {\bf 107}, 503 (2009).

\bibitem{msa} J.S.~H\o ye {\it et al.}, Mol. Phys. {\bf 32}, 209 (1976).

\bibitem{bmw} J.-P. Blaizot, R. M\'endez-Galain, and N. Wschebor, Phys. Lett.
B {\bf 632}, 571  2006.

\bibitem{bmw2} F. Benitez {\it et al.}, \pre {\bf 80}, 030103(R) (2009). 

\bibitem{scoza}
D.~Pini, G.Stell and N.B.Wilding,  Mol. Phys. {\bf 95}, 483 (1998).

\bibitem{scoza2}
J.S.H\o ye, D.Pini and G.Stell, Physica A {\bf 279}, 213 (2000).

\bibitem{reinerhoye}
A.Reiner and J.S.H\o ye, \pre {\bf 72}, 061112 (2005). 

\bibitem{field}
A.~Pelissetto and E.~Vicari, Phys. Rep. {\bf 368}, 549 (2002).

\bibitem{max1} 
A.~Parola, D.Pini and L. Reatto, Phys. Rev. E {\bf 48}, 3321 (1993).

\bibitem{max2} 
C.D.~Ionescu, A.~Parola, D.Pini and L. Reatto, Phys. Rev. E {\bf 76}, 031113 (2007).

\bibitem{wet2}
N. Tetradis and C. Wetterich, Nucl. Phys. B {\bf 383}, 197 (1992).

\bibitem{prl}
A. Parola, D. Pini, L. Reatto, Phys. Rev. Lett {\bf 100}, 165704 (2008).  

\bibitem{molphys}
A. Parola, D. Pini, L. Reatto, A. Parola, D. Pini, L. Reatto; Mol. Phys.  {\bf 107}, 503 (2009).

\bibitem{litim}
D. Litim, Int. J. Mod. Phys. A, {\bf 16}, 2081 (2001). 

\bibitem{catania} 
A. Bonanno and G. Lacagnina, Nucl. Phys. B {\bf 693}, 36 (2004)

\bibitem{caillol2} 
J.M. Caillol, ``The non-perturbative renormalization group in the ordered phase" 
http://xxx.lanl.gov/abs/1109.4024v2

\bibitem{drop}
M.E. Fisher, Physics (N.Y.) {\bf 3}, 255 (1967);
W. Klein, \prb {\bf 21}, 5254 (1980).

\bibitem{meroni}
A.Parola, A.Meroni and L.Reatto, \prl {\bf  62}, 2981, (1989).

\bibitem{ising}
D. Pini, A. Parola, L. Reatto, J. Stat. Phys. {\bf 72}, 1179 (1993).

\bibitem{barocchi}
F.Barocchi {\it et al.}, J. Phys. Condens. Matter {\bf 9}, 8849 (1997). 

\bibitem{tau}
M.Tau, A.Parola, D.Pini and L.Reatto \pre {\bf 52}, 2644 (1995).

\bibitem{rpm}
A.Parola and D.Pini, Mol. Phys. {\bf 109}, 2989 (2011). 

\bibitem{reiner1} A.Reiner and G.Kahl, \pre {\bf 65}, 046701 (2002).

\bibitem{reiner2} A.Reiner, J.Stat.Phys. {\bf 118}, 1107, (2005); 
A.Reiner and G.Kahl,  J.Stat.Phys. {\bf 118}, 1129 (2005).

\bibitem{vink} 
F.Lo Verso, R.L.C.Vink, D.Pini and L.Reatto, \pre {\bf 73}, 061407 (2006). 

\bibitem{loverso2}
D.Pini, F.Lo Verso, M.Tau, A.Parola and L.Reatto, \prl {\bf 100}, 055703 (2008).

\bibitem{expcin}
M.Seul and D.Andelman, Science, {\bf 267}, 476 (1995).

\bibitem{imperio}
A.Imperio and L.Reatto, J. Phys. Condens. Matter, {\bf 16}, S3769 (2004);
\pre {\bf 76}, 040402 (2007). 

\bibitem{cina}
D.Pini, Ge Jialin, A.Parola and L.Reatto, Chem. Phys. Lett. {\bf 327}, 209 (2000).

\bibitem{likos}
C.N.Likos, H.L\"owen, M.Watzlawek, O.Abbas Jucknischke, J.Allgaier and D.Richter,
\prl {\bf 80} 4450 (1998).  

\bibitem{camargo}
M.Carmargo and C.N.Likos, \prl {\bf 104}, 078301 (2010). 

\bibitem{loverso}
F.Lo Verso, M.Tau and L.Reatto, J. Phys. Condens. Matter {\bf 15}, 1505 (2003).

\bibitem{brognara1} A.Brognara, A.Parola and L.Reatto, \pre {\bf 64}, 026122 (2001).

\bibitem{brognara2} A.Brognara, A.Parola and L.Reatto, \pre {\bf 65}, 066113 (2002).

\bibitem{dickman} 
R.Dickman and G.Stell, in {\it Simulation and Theory of Electrostatic
Interactions in Solutions} edited by L. R. Pratt and G.  Hummer (AIP, Woodbury, NY, 1999).

\bibitem{panagio} A.Z.Panagiotopoulos and S.K.Kumar, Phys. Rev. Lett. {\bf 83}, 2981 (1999).

\bibitem{mix} A.Parola and L.Reatto, \pra 6600, (1991).

\bibitem{phenom} M.E.Fisher, Phys. Rev. {\bf 176}, 257 (1968); R.B.Griffiths and 
J.C.Wheeler \pra {\bf 2}, 1047 (1970). 

\bibitem{crox}D.Pini,A.Parola and L.Reatto, J. Phys. Condensed Matter {\bf 9}, 1417 (1997).

\bibitem{bathia} A.B.Bathia and D.E.Thornton, \prb {\bf 2}, 3004 (1970).

\bibitem{mixlj} D.Pini, A.Parola and L.Reatto, J. Stat. Phys. {\bf 100}, 13 (2000).

\bibitem{mixyuk} D.Pini, M.Tau, A.Parola and L.Reatto, \pre {\bf 67}, 046116 (2003).

\bibitem{isfl} M.J.P.Nijmeijer, A.Parola, L.Reatto, Phys. Rev. E {\bf 57}, 465 (1998).

\bibitem{shouten} J.A.Schouten, Phys. Rep. {\bf 172}, 35 (1989).

\bibitem{arkr} J.A.Schouten, A.Deerenberg and N.J.Trappenier, Physica {\bf 81A}, 151 (1975).

\bibitem{nekr} N.J.Trappenier and J.A.Schouten, Physica {\bf 73}, 546 (1974). 

\bibitem{gianinetti} P.Gianinetti and A.Parola \prb {\bf 63}, 104414 (2001). 

\bibitem{holm} C.Holm and W.Janke, \prb {\bf 48}, 936 (1993). 

\bibitem{sandvik} A.W.Sandvik, \prl {\bf 80}, 5196 (1998).

\bibitem{chn} S.Chakravarty, B.I.Halperin, and D.R.Nelson, \prb {\bf 39},2344 (1989).

\bibitem{gianinetti2} P.Gianinetti, A.Parola and L.Reatto, 
in {\sl Advances in Quantum Many-Body Theory Vol. 5: 150 years of Quantum Many-Body Theory},
R.F. Bishop, K.A. Gernoth, N.R. Walet Eds. World Scientific, Singapore, (2001).

\bibitem{spanu} L.Spanu and A.Parola \prb {\bf 72}, 174418 (2005). 

\bibitem{dupuis2} A.Ran\c con and N.Dupuis, \prb {\bf 83}, 172501 (2011). 

\bibitem{tarjus} G.Tarjus, M.L.Rosinberg, E.Kierlik and M.Tissier, Mol.Phys. {\bf 109}, 2863 (2011). 

\bibitem{verlet} L.Verlet and J.J.Weis, \pra {\bf 5}, 939 (1972). 

\end{thebibliography}
\end{document}